\documentclass[12pt]{article}
\usepackage{geometry}
\usepackage{graphicx} % Required for inserting images
\usepackage{setspace}
\usepackage{changepage}
\usepackage{microtype}
\usepackage{subfigure}
\usepackage{graphicx}
\usepackage{array}
\usepackage{bm}
\usepackage{wrapfig}
\usepackage{float}
\usepackage{authblk}
\usepackage{xcolor} 
\usepackage{colortbl}

\include{macros}

\definecolor{darkmagenta}{rgb}{0.55, 0, 0.55}
\title{Inference for Stable Variable Importance across Multi-source Domains via Adversarial Learning}

\author[1]{Zitao Wang}
\author[2]{Nian Si}
\author[3]{Zijian Guo}
\author[4]{Molei Liu\footnote{Email: ml4890@cumc.columbia.edu}}

\affil[1]{Department of Statistics, Columbia University.}
\affil[2]{Department of Industrial Engineering and Decision Analytics, Hong Kong University of Science and Technology.}
\affil[3]{Department of Statistics, Rutgers University.}  
\affil[4]{Department of Biostatistics, Columbia Mailman School of Public Health.}

\date{}

\begin{document}

\setstretch{1.1} 
\maketitle

\begin{abstract}

%As part of enhancing the interpretability of machine learning, it is of renewed interest to quantify and infer the predictive importance of certain exposure covariates. Modern scientific studies often collect data from multiple sources with distributional heterogeneity. Thus, measuring and inferring stable associations across multiple environments is crucial in reliable and generalizable decision-making. In this paper, we propose MIMAL, a novel statistical framework for \textbf{M}ulti-source stable \textbf{I}mportance \textbf{M}easure via \textbf{A}dversarial \textbf{L}earning. MIMAL measures the importance of some exposure variables by maximizing the worst-case predictive reward over the source mixture. Our framework allows various machine learning methods for confounding adjustment and exposure effect characterization. For inferential analysis, the asymptotic normality of our introduced statistic is established under a general machine learning framework that requires no stronger learning accuracy conditions than those for single source variable importance. Numerical studies with various types of data generation setups and machine learning implementation are conducted to justify the finite-sample performance of MIMAL. We also illustrate our method through a real-world study of Beijing air pollution in multiple locations.

The quantification and inference of predictive importance for exposure covariates have recently gained significant attention in the context of interpretable machine learning. Contemporary scientific investigations often involve data originating from multiple sources with distributional heterogeneity. It is imperative to introduce a new notation of the variable importance measure that is stable across diverse environments. In this paper, we introduce MIMAL (Multi-source Importance Measure via Adversarial Learning), a novel statistical framework designed to quantify the importance of exposure variables by maximizing the worst-case predictive reward across source mixtures. The proposed framework is adaptable to a broad spectrum of machine learning methodologies for both confounding adjustment and exposure effect characterization. We establish the asymptotic normality of the data-dependent estimator of the multi-source variable importance measure under a general machine learning framework. Our framework requires the similar learning accuracy conditions compared to those required for single-source variable importance analysis. The finite-sample performance of MIMAL is demonstrated through extensive numerical studies encompassing diverse data generation scenarios and machine learning implementations. Furthermore, we illustrate the practical utility of our approach in a real-world case study of air pollution in Beijing, analyzing data collected from multiple locations.

\vspace{0.3cm}
\noindent \textbf{Keywords:} Nonparametric variable importance; Stable association; Group distributionally robust learning; Statistical inference with machine learning; Interpretable machine learning. 

\end{abstract}

\section{Introduction}

\subsection{Background and motivation}

In scientific studies, explainability and interpretability of the regression analysis are crucial in making scientific decisions. Despite the prevalence of addressing prediction problems with modern machine learning (ML) models such as random forests, kernel machine and neural networks, the interpretability of general ML methods is an as important yet under-explored problem. To measure the nonparametric variable importance of some exposure $X$ on some outcome $Y$ adjusting for $Z$, one commonly used strategy is the so-called Leave Out COvariates (LOCO) method \citep[e.g.]{nonparametric-vi,tansey2022holdout}. The main idea of LOCO is to fit two (nested) ML models separately for $Y\sim Z$ and $Y\sim (X,Z)$ and contrast their prediction loss characterized by certain evaluation metric such as the mean squared prediction error (MSPE). Intuitively, a large reduction on the prediction loss when including $X$ can indicate a strong variable importance of $X$ on $Y$ given $Z$.

Nonetheless, in broad fields such as biomedical research, it is common to integrate data collected from multiple heterogeneous sources or populations for integrative regression analyses. In this situation, it is of great interest to capture important covariates displaying similar or stable effects on $Y$ across different sources. For example, some recent genome-wide and phenome-wide association studies like \cite{verma2024diversity} focused on finding stable genotype-phenotype associations across different ethnicity groups or institutions. By ``stable'' association between $Y$ and $X$ (given $Z$), one requires not only their dependence to exist across all sources but also the direction or pattern of $X$'s effect on $Y$ to be shared by the sources. Scientifically, such relationships tend to be more generalizable to new environments.

\subsection{Related works}
Before introducing our main results and contributions, we shall review two lines of research that are closed relevant to our work.

\vspace{-0.25cm}

\paragraph{ML-agnostic variable importance.} Classic inference procedures based on parametric linear or generalized linear models suffer from their limited capacity in capturing more complex effects of $X$ on $Y$ as well as adjusting for high-dimensional confounders $Z$. In recent literature, there arises a great interest in characterizing and inferring variable importance based on general ML algorithms such as random forests or neural networks. In specific, \cite{nonparametric-vi} established the asymptotic normality and inference approach for the LOCO $R^2$-statistic constructed with general ML. \cite{general-framework} extended this to a more general framework accommodating generic importance assessment functions and developed semiparametric efficient estimation. \cite{floodgate} addressed the inference of LOCO under the model-X framework with the knowledge of $\mathbb{P}(X\mid Z)$. To improve the power of LOCO, \cite{williamson2020efficient} and others studied Shapely value as an alternative strategy less prone to high correlation among the covariates; and \cite{verdinelli2024decorrelated} further developed a decorrelated variable importance measure framework attaining better performance. However, to our best knowledge, none of existing methods in this track can be used to infer important variables holding a similar or stable relationship with $Y$ across multi-source heterogeneous data sets. 

\vspace{-0.25cm}

\paragraph{Group Distributionally Robust Learning.} To characterize multi-source generalizable effects, our framework will be based on an adversarial learning construction relevant to the maximin regression and group distributionally robust learning (DRoL) widely studied in recent years \citep[e.g.]{mohri2019agnostic}. With multi-source data, group DRoL aims to optimize the worst-case performance of ML models on all sources. In this framework, \cite{GroupDRO} proposed avoiding over-fitting of over-parameterized neural networks with stronger regularization, which largely improves the worst-group accuracy. \cite{maximin-effect} developed a maximin regression framework to enhance the generalizability of linear models by maximizing their smallest reduced variance on multiple sources. \cite{Zijian_groupDRO} extended this maximin method to accommodate general ML estimation with the least square loss. \cite{zhang2024optimal} derive the optimal online learning strategy and sample size for group DRoL under general models. \cite{mo2024minimax}, \cite{zhan2024transfer}, and other recent works aimed at fixing the over-conservative issue of group DRoL using different strategies to guide the adversarial learning procedure. Nevertheless, most existing works in group DRoL focus on its optimization and prediction, with little attention paid to statistical inference and interpretation. We note that \cite{dominik} and \cite{Zijian_maxmininfer} studied the inference of maximin effects in linear parametric models, which is insufficient for ML-agnostic variable importance assessment. 

\subsection{Our results and contributions}
Despite the importance, there is a large lack of framework for measuring stable and generalizable variable importance shared across multiple sources, together with the related statistical inference tools based on general machine learning methods, which limits the interpretability of ML on multi-source data. To fill this gap, we propose a novel inferential framework for Multi-source stable variable Importance Measure via Adversarial Learning (MIMAL). In this framework, we introduce a non-parametric variable importance measure named as MIMAL that characterizes the stable and generalizable dependence between $Y$ and $X$ across multiple heterogeneous populations while allowing each source to adjust for the confounding $Z$ freely. We further introduce to a group adversarial learning task that trains a unified ML model to optimize the minimum predictive power of $X$ across all convex mixtures of the sources, and then infer the variable importance (MIMAL) as the objective value of this task. Importantly, we incorporate advanced optimization tools to enable the use of complex ML methods like neural networks and gradient boosting. This ensures the MIMAL framework is flexible and accommodates a wide range of ML methods as well as general variable importance criteria such as log-likelihood.

Asymptotic unbiasedness and normality are established for our empirical estimator of MIMAL, with a key assumption on the $o(n^{-1/4})$-convergence of the ML estimators in the typical regression task on every specific mixture of the sources. Interestingly, such a rate requirement is not stronger than those used for the ML-agnostic variable importance inference on a single homogeneous population \citep[e.g.]{general-framework}.  Our theoretical justifications are more technically involved than existing theory for distributionally robust stochastic programming \citep{shapiro}, with the main complication lying in the slow convergence rate and blackbox nature of general and complex ML methods. To handle this challenge, we establish Neyman orthogonality of the objective value function as well as $o(n^{-1/4})$-convergence of the nuisance ML and weight estimators in group adversarial learning, which in turn, facilitates the automatic elimination of excessive bias from ML. This result can be viewed as a novel extension of existing asymptotic theory for ML-agnostic variable importance \citep[Theorem 1]{general-framework} from the single-source regression setting to adversarial optimization problems. Particularly, the single-source variable importance is based on minimization problems while our multi-source version is based on minimax optimization problem, where the Neyman orthogonality requires more effort to establish. Numerical studies with various data types and ML architectures demonstrate good finite-sample performances of our method in terms of statistical inference and support our theoretical establishments. 

\subsection{Outline of the Paper}
The rest of this paper is organized as follows. In Section \ref{sec:framework}, we introduce the problem setup and the stable variable importance measure across multiple sources. In Section \ref{sec:method}, we propose the empirical estimation and inference approach for MIMAL. In Section \ref{sec:thm}, we establish the theoretical framework for MIMAL and justify the asymptotic normality of our proposed estimator. In Section \ref{sec:sim}, we conduct simulation studies with various setups to demonstrate the finite-sample inferential performance of our method. In Section \ref{real-data}, we illustrate our method with application to learn the variable importance for air pollution in Beijing. In Section \ref{sec:disc}, we conclude our paper with some discussion on future directions. 

\section{Setup and Framework}\label{sec:framework}
Denote by $[n]=\{1,2\ldots,n\}$ for any positive integer $n$. Suppose there are $M$ heterogeneous source populations with outcome $Y\supm$, exposure variables $X\supm\in\mathcal{X}$, and adjustment covariates $Z\supm\in\mathcal{Z}$ generated from the probability distribution $\mathbb{P}\supm_{Y|X,Z}\times \mathbb{P}\supm_{X,Z}$ for each source $m\in[M]$. We use the lowercase $(y\supm_i,x\supm_i,z\supm_i)$ for $i\in[n_m]$ to represent the $n_m$ observations on source $m$ and denote by $\bold{y}\supm,\bold{X}\supm$, $\bold{Z}\supm=\{y\supm_i:i\in[n_m]\}, \{x\supm_i:i\in[n_m]\}$, $\{z\supm_i:i\in[n_m]\}$, $\bold{D}\supm=\{\bold{y}\supm,\bold{X}\supm,\bold{Z}\supm\}$, and $\bold{D}=\{\bold{D}\supm:m\in[M]\}$. Also, let $\mathbb{E}\supm$ and $\widehat{\mathbb{E}}\supm$ respectively denote the population expectation and sample mean operators (i.e., $\widehat{\mathbb{E}}\supm=n_m^{-1}\sum_{i=1}^{n_m}$) on source $m$. Let $L^2(\mathcal{X},\mathcal{Z})$ and $L^2(\mathcal{Z})$ respectively denote the space of square-integrable functions of $(X,Z)$ and $Z$ with respect to all $\mathbb{P}\supm_{X,Z}$ for $m\in[M]$. We use  $\diag(\boldsymbol{a})$ to represent the diagonal matrix of any vector $\boldsymbol{a}$. The capital letter $\Phi(\cdot)$ denotes the cumulative distribution function of the standard normal distribution.

To facilitate the discussion, we first introduce in Section \ref{sec:setup:noZ} the multi-source stable variable importance in a simplified setup without adjustment covariates $Z$, and then extend it in Section \ref{sec:method:adj} to a more general setting with source-specific adjustment on $Z$.

\subsection{The simplified setup without $Z$}\label{sec:setup:noZ}

To regress $Y$ against $X$ (denoted as $Y\sim X$) on the $m$-th source, we introduce an objective function $\ell\{Y, f(X)\}$, where $f(\cdot)$ is a prediction function capturing the effect of $X$ on $Y$, and $\ell(\cdot): \mathbb{R}^2 \to \mathbb{R}$ measures its goodness-of-fit. We shall consider that $f(\cdot)$ belongs to a function class $\mathcal{F}$ that will be specified in the following discussion. For a continuous outcome variable, a common choice for $\ell$ is the $R^2$ coefficient or the negative squared error: $\ell(y, u) = -(y - u)^2$ with $f(X)$ intended to characterize $\mathbb{E}[Y \mid X]$. For a binary outcome variable, one can use the logistic log-likelihood $\ell(y,u)=yu-\log(1+e^{u})$ with $f(X)$ capturing ${\rm logit}\{{\rm P}(Y=1\mid X)\}$ or the negative hinge loss. To quantify the predictive importance of $X$ in predicting $Y$ on each source $m$, we introduce the reward function 
\[
R\supm(f):={\mathbb{E}}\supm \big[\ell\{Y,{f}(X)\}-\ell(Y,0)\big],
\] 
and its optimal value $I\supm_X:=\max_{f\in\mathcal{F}} R\supm(f)$. A larger optimal reward $I\supm_X$ indicates that the variable $X$ has a stronger importance in predicting $Y$. 

Then, to measure the multi-source stable effect of $X$ on $Y$, we define the adversarial reward function as the minimum of $R\supm(f)$ over all $M$ sources, i.e., $R_{\rm adv}(f) \coloneqq\min_{m\in[M]}R\supm(f)$. Based on this, we define the multi-source (MIMAL) variable importance as
\[
I_X^*\coloneqq \max_{f\in\mathcal{F}} R_{\rm adv}(f) =\max_{f\in\mathcal{F}}\min_{m\in[M]}{\mathbb{E}}\supm \big[\ell\{Y,{f}(X)\}-\ell(Y,0)\big].
\]
The adversarial reward $R_{\rm adv}(f)$ characterizes the minimum gain across the $M$ sources by predicting the outcome with $f(X)$ compared to ignoring the prediction power of $X$. The optimal value $I^*_X$ provides a measure of variable importance of $X$ that is shared across multiple sources. We shall refer to this measure as the stable variable importance across $M$ sources. Compared to the single-source $I\supm_X$, $I^*_X$ provides a more conservative assessment of $X$'s predictiveness as seen from the fact $I^*_X\leq I\supm_X$ for every $m\in[M]$. Consequently, $I^*_X$ will be zero when $Y$ is independent of $X$ and $I\supm_X=0$ on any of the sources $m$.

\subsection{Incorporating confounding adjustments}\label{sec:method:adj}

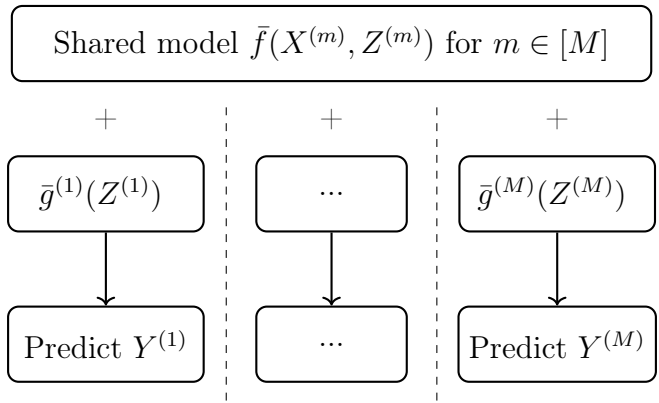
\begin{figure}[htbp]
    \centering
\begin{tikzpicture}
\node[draw, rounded corners, thick, minimum width=8.5cm, minimum height=1cm, align=center] (shared) at (0,0) {Shared model $\bar{f}(X\supm,Z\supm)$ for $m\in[M]$};
\node[draw, rounded corners, thick, minimum width=2.6cm, minimum height=1cm, align=center] (specific1) at (-3,-2) { $\bar{g}^{(1)}(Z^{(1)})$ };
\node[draw, rounded corners, thick, minimum width=2cm, minimum height=1cm, align=center] (specific2) at (0,-2) {...};
\node[draw, rounded corners, thick, minimum width=2.6cm, minimum height=1cm, align=center] (specific3) at (3,-2) { $\bar{g}^{(M)}(Z^{(M)})$ };
\node[thick] at (-3,-1) {+};
\node[thick] at (3,-1) {+};
\node[thick] at (0,-1) {+};

% Add dashed vertical lines
\draw[dashed] (-1.4,-4.75) -- (-1.4,-0.75);
\draw[dashed] (1.4,-4.75) -- (1.4,-0.75);

% Add circular nodes below each specific node with a consistent size
\node[draw, rounded corners, thick, minimum width=2.6cm, minimum height=1cm, align=center] (circle1) at (-3,-4) {Predict $Y^{(1)}$};
\node[draw, rounded corners, thick, minimum width=2cm, minimum height=1cm, align=center] (circle2) at (0,-4) {...};
\node[draw,rounded corners, thick, minimum width=2.6cm, minimum height=1cm, align=center] (circle3) at (3,-4) {Predict $Y^{(M)}$};

% Draw arrows from specific nodes to circular nodes
\draw[->, thick] (specific1) -- (circle1);
\draw[->, thick] (specific2) -- (circle2);
\draw[->, thick] (specific3) -- (circle3);

\end{tikzpicture}
\caption{ \label{fig:structure} Model structure encoded in the MIMAL objective function (\ref{equ:Ix}).}
\end{figure}

In the presence of confounding covariates $Z$, for the $m$-th source data, we introduce two models $b\supm(\cdot)$ and $g\supm(\cdot)$ for $Z$, which are assumed to  belong to the function class $\mathcal{G}\supm$ that are specified in the prelude to Theorem \ref{prop:obj}. The LOCO variable importance of $X$ in each single-source can be defined as $I\supm_X:=\max_{f\in\mathcal{F},g\supm\in\mathcal{G}\supm}R\supm(f,g\supm)$, with the reward
\begin{equation}
R\supm(f,g\supm):=\mathbb{E}\supm\ell\{Y,f(X,Z)+g\supm(Z)\}-\mathbb{E}\supm\ell\{Y,\bar{b}\supm(Z)\},   
\label{equ:single:vi}
\end{equation}
where the baseline regression model $\bar{b}\supm$ for $Y\supm\sim Z\supm$ is defined as
\begin{equation}
\bar{b}\supm=\argmax_{b\supm\in\mathcal{G}}\mathbb{E}\supm\ell\{Y,b\supm(Z)\}.
\label{equ:b}
\end{equation}
Note that the  $f(\cdot)\in \mathcal{F}$ in Equation \eqref{equ:single:vi} encodes the effect of $X$ on $Y$ shared by all the sources and allowed to interact with $Z$, and $g\supm(\cdot)\in \mathcal{G}\supm$ is the source-specific adjustment function of $Z$. As illustrated in Figure \ref{fig:structure}, the prediction model for $Y\sim (X,Z)$ on source $m$ is supposed to be the combination of these two components. Thus, $I\supm_X$ characterizes the importance of $X$ on $Y$ given $Z$ by contrasting the prediction performances between the two model classes $f(X,Z)+g\supm(Z)$ and $b\supm(Z)$, with the former including $(X,Z)$ while the latter only including $Z$. {Note that $g\supm(Z)$ and $b\supm (Z)$ share the same space $\mathcal{G}\supm$, which ensures the model space of $f(X,Z)+g\supm(Z)$ contains that of $b\supm(Z)$, i.e., two models being nested and, thus, $I_X\supm\geq 0$.}

\begin{remark}
For concrete applications, we usually impose certain structural assumptions on $f$. As an example, one could simply take $f(X,Z)=f_1(X)$ to capture the stable effect solely from $X$ without involving any interactions with $Z$. In clinical studies with $Y$ being some disease outcome and $Z\in\{0,1\}$ for a treatment indication, one may take $f(X,Z)=Zf_2(X)$ with $f_2(X)$ characterizing the individual treatment effect determined by $X$. This could be used to decide if $X$ is an important effect modifier.     
\end{remark}

Denote by $\boldsymbol{g}=\{g\supm:m\in[M]\}$ and $\boldsymbol{\mathcal{G}}=\mathcal{G}^{(1)}\times\ldots,\times\mathcal{G}^{(M)}$. Similar to the maximin construction introduced in Section \ref{sec:setup:noZ}, we define the adversarial reward as:
\begin{equation}
R_{\rm adv}(f,\boldsymbol{g})\coloneqq\min_{m\in[M]}R\supm(f,g\supm),
\label{equ:R}    
\end{equation}
and the MIMAL variable importance measure (with adjustments on $Z$) as
\begin{equation}
I^*_X:=R(\bar{f},\bar{\boldsymbol{g}}),\quad \mbox{where}\quad(\bar{f},\bar{\boldsymbol{g}})\in\argmax_{f\in \mathcal{F},\boldsymbol{g}\in{\boldsymbol{\mathcal{G}}}} R_{\rm adv}(f,\boldsymbol{g}).
\label{equ:Ix}
\end{equation}
Distinguished from the setup in Section \ref{sec:setup:noZ}, our formulation here includes a source-specific component $g\supm(Z)$ to be fitted together with $f$ in (\ref{equ:Ix}) and freely changed across the sources, which allows for confounding adjustments heterogeneously across the sources. When $X$ is predictive of $Y$ given $Z$, the optimal $\bar{g}\supm$ is typically unequal to the baseline $\bar{b}\supm$ fitted without $X$. Without this $g\supm$ component, $f(X,Z)$ alone would become non-nested with the heterogeneous $\bar{b}\supm$'s and there would be no more guarantee that $I^*_X\geq 0$.

\begin{remark}
Usually, $\bar{f}$ and $\bar{\boldsymbol{g}}$ defined in (\ref{equ:Ix}) cannot be separately identifiable. For example, $\bar{f}+c$ and $\bar{\boldsymbol{g}}-c$ for any constant $c\neq 0$ is also the maximizer of $R_{\rm adv}(f,\boldsymbol{g})$ as long as they belong to $\mathcal{F}$ and $\boldsymbol{\mathcal{G}}$. Nevertheless, this will not cause any essential identification issues to $I^*_X$ of our interests. We can show they are still uniquely identifiable under the strict concavity Assumption \ref{A2a} tending to hold in general. For the empirical estimation of $I^*_X$, it is helpful to achieve separate identifiability on the {\em estimators} of $\bar{f}$ and $\bar{\boldsymbol{g}}$ to ensure proper convergence of the learners. For example, for parametric regression, one could set $g\supm(Z)=\gamma\supm_0+Z\trans\gamma\supm$ and $f(X,Z)=X\trans\beta+(X\otimes Z)\trans\theta$ where $X\otimes Z$ consists of interaction terms between $X$ and $Z$. In this case, $g\supm$ and $f$ do not share any basis so they can be separately identified and estimated. Similar strategies are used in our construction for more complex ML methods like neural networks; see Appendix.

\label{rem:1}
\end{remark}

\section{Method}\label{sec:method}
\subsection{Adversarial learning and inferential framework}

In this section, we describe our empirical estimation and inference approach for $I^*_X$. To begin with, we introduce an equivalent form of the minimum reward objective function in (\ref{equ:R}) as
\begin{equation}
R(q,f,\boldsymbol{g};\boldsymbol{b}):=\sum_{m=1}^Mq_m\left[\mathbb{E}\supm\ell\{Y,f(X,Z)+g\supm(Z)\}-\mathbb{E}\supm\ell\{Y,b\supm(Z)\}\right],
\label{equ:pop:maxmin}
\end{equation}
where $\boldsymbol{b}=\{b\supm:m\in[M]\}$, and $q=(q_1,\ldots,q_M)\trans$ is a set of probabilistic weights for the sources defined on the simplex $\Delta^M$. Noting that $R_{\rm adv}(f,\boldsymbol{g})=\min_{q\in\Delta^M}R(q,f,\boldsymbol{g};\bar{\boldsymbol{b}})$ where $\bar{\boldsymbol{b}}=\{\bar{b}\supm:m\in[M]\}$, we have
\begin{equation}
I^*_X:=\max_{f\in \mathcal{F},\boldsymbol{g}\in\boldsymbol{\mathcal{G}}}R_{\rm adv}(f,\boldsymbol{g})=\max_{f\in \mathcal{F},\boldsymbol{g}\in\boldsymbol{\mathcal{G}}}\min_{q\in\Delta^M}R(q,f,\boldsymbol{g};\bar{\boldsymbol{b}}).
\label{obj_q}
\end{equation}
In this way, (\ref{equ:Ix}) is converted to a group adversarial learning problem on the right hand side of (\ref{obj_q}) with learners $f$ and $\boldsymbol{g}$ and adversarial weights in $q$. In Theorem \ref{prop:obj}, we establish an identification strategy for $I^*_X$ as well as its corresponding arguments $(\bar{q},\bar{f},\bar{\boldsymbol{g}})$. In this treatment, we take $\mathcal{F}$ and $\boldsymbol{\mathcal{G}}$ to be norm-bounded subset of the space of square-integrable functions, e.g take $\mathcal{F} \coloneqq \{f\in L^2\{\mathcal{X},\mathcal{Z}\} | \mathbb{E}_{X,Z}(f^2) \leq K\}$ for some large $K>0$, and similarly define each component of the product space $\boldsymbol{\mathcal{G}} = \mathcal{G}^{(1)}\times\ldots\times \mathcal{G}^{(M)}$. The candidate function class $\mathcal{F}$ and $\boldsymbol{\mathcal{G}}$ are now convex, closed and bounded.

\begin{theorem}
\label{prop:obj} 
Assume that $\mathbb{E}\supm \ell\{Y,\,\cdot\,\}$ is concave on the function class $L^2(\mathcal{X},\mathcal{Z})$ and recall that $\bar{\boldsymbol{b}}$ is as defined in (\ref{equ:b}), then the objective $\max_{f\in \mathcal{F},\boldsymbol{g}\in\boldsymbol{\mathcal{G}}}\min_{q\in\Delta^M}R(q,f,\boldsymbol{g};\bar{\boldsymbol{b}})$ has a non-negative and finite optimal value $I^*_X=R(\bar{q},\bar{f},\bar{\boldsymbol{g}};\bar{\boldsymbol{b}})$ achieved at the population saddle point,
\begin{equation}
\bar{q}\in \argmin_{q\in\Delta^M} \max_{f\in\mathcal{F},\boldsymbol{g}\in\boldsymbol{\mathcal{G}}}R(q,f,\boldsymbol{g};\bar{\boldsymbol{b}}),\quad\mbox{and}\quad (\bar{f},\bar{\boldsymbol{g}})\in\argmax_{f\in\mathcal{F},\boldsymbol{g}\in\boldsymbol{\mathcal{G}}} \min_{q\in\Delta^M} R(q,f,\boldsymbol{g};\bar{\boldsymbol{b}}).
\label{equ:nash:pop}
\end{equation}
\end{theorem}

\begin{remark}
The constraint in problem (\ref{equ:nash:pop}) is actually a learning task to obtain the optimal prediction model $(f,\boldsymbol{g})$ on the mixture of sources with any fixed set of probabilistic weights in $q$. Empirically, this is supposed to be realized through ML techniques. Intuitively, (\ref{equ:nash:pop}) finds the optimal adversarial weights by looking into the rewards $R(q,f,\boldsymbol{g};\bar{\boldsymbol{b}})$ for all $q\in\Delta\supm$ and picking the smallest one from them. Under the concavity assumption in Theorem \ref{prop:obj}, the validity of this strategy is implied by Sion's minimax theorem \citep{sion} that
\[\max_{f\in \mathcal{F},\boldsymbol{g}\in\boldsymbol{\mathcal{G}}}\min_{q\in\Delta^M}R(q,f,\boldsymbol{g};\bar{\boldsymbol{b}})=\min_{q\in\Delta^M}\max_{f\in \mathcal{F},\boldsymbol{g}\in\boldsymbol{\mathcal{G}}}R(q,f,\boldsymbol{g};\bar{\boldsymbol{b}}).
\]
This objective can be viewed as a zero-sum game between two ``players'' $(f,\boldsymbol{g})$ and $q$ in the sense that increasing the reward $R$ in favor of $(f,\boldsymbol{g})$ will incur a loss with respect to $q$ and vice versa. Correspondingly, the saddle point $(\bar{q},\bar{f},\bar{\boldsymbol{g}})$ is also called the Nash equilibrium to this game because at this point, no player has the incentive to unilaterally deviate.
\end{remark}

There are two main advantages of transforming (\ref{equ:Ix}) to the group adversarial learning problem in (\ref{obj_q}), both of which can be seen from our learning algorithms to be introduced next. First, as seen from Theorem \ref{prop:obj}, we replace the discrete minimum function $R_{\rm adv}(f,\boldsymbol{g})$ with $R(q,f,\boldsymbol{g};\bar{\boldsymbol{b}})$ that is continuous in both $q$ and $(f,\boldsymbol{g})$ and, thus, is easier to optimize in practice with gradient-based ML methods to be introduced later. Second, as will be established in Section \ref{sec:thm}, the saddle point $(\bar{q},\bar{f},\bar{\boldsymbol{g}})$ plays an central role in characterizing the asymptotic distribution of the empirical estimation for $I_X^*$ so their estimations are required to facilitate statistical inference on $I^*_X$.

Based on the aforementioned formulation, we propose the MIMAL approach for the point and interval estimation of $I^*_X$ outlined in Algorithm \ref{algo1}. It contains two regression steps including (i) learning the baseline model for $Y\supm\sim Z\supm$ in each source $m$; and (ii) a sample-level group adversarial learning motivated by Theorem \ref{prop:obj}. Step (i) is a standard regression task allowing flexible use of general ML tools. Step (ii) is the empirical version of (\ref{equ:nash:pop}) and more complicated to solve. We introduce a gradient-based optimization procedure in Section \ref{sec:method:grad} that jointly updates $q$ and $f,\boldsymbol{g}$ to attain the saddle-point solution. It facilitates the use of a broad class of ML methods such as neural networks and gradient boosting. In both steps, we adopt $K$-fold cross-fitting to avoid the over-fitting bias caused by complex ML methods, which is in a similar spirit with \cite{doubleML} and others.

\begin{algorithm}[H]
\SetAlgoLined
\textbf{Input:} Multi-source sample observations $\bold{D}\supm$ for $m\in[M]$. Pre-specified ML model spaces $\tilde{\mathcal{F}}_{\lambda}$ and $\tilde{\boldsymbol{\mathcal{G}}}_{\lambda}=\tilde{\mathcal{G}}^{(1)}_{\lambda}\times\ldots\times\tilde{\mathcal{G}}^{(M)}_{\lambda}$ with tuning parameters in $\lambda$.
 
\For{$m = 1,\ldots,M$}{
Randomly split $\bold{D}\supm$ into $K\geq 2$ equal-sized folds $\{\bold{D}\supm_{[k]}:k\in[K]\}$. Let $\bold{D}_{[\textit{-}k]}\supm=\bold{D}\supm\setminus \bold{D}_{[k]}\supm$, and $\widehat{\mathbb{E}}_{[k]}\supm$ and $\widehat{\mathbb{E}}_{[\textit{-}k]}\supm$ be the empirical mean operators on $\bold{D}_{[k]}\supm$ and $\bold{D}_{[\textit{-}k]}\supm$ respectively.

 }
 \For{$k = 1,\ldots,K$}{
 {\bf (i)} Learn the baseline models $\widehat{{\boldsymbol{b}}}_{[\textit{-}k]}=\big\{\widehat{b}\supm_{[\textit{-}k]}:m\in[M]\big\}$ where 
 \[
 \widehat{b}\supm_{[\textit{-}k]} = \argmax_{b\supm\in\tilde{\mathcal{G}}_{\lambda}\supm}\widehat{\mathbb{E}}_{[\textit{-}k]}\supm\ell\{Y,b\supm(Z)\}.
 \]
 
 {\bf (ii)} Solve the maximin optimization problem 
 \begin{align}
 \widehat{q}_{[\textit{-}k]}&= \argmin_{q\in\Delta^M}\max_{f\in \tilde{\mathcal{F}}_{\lambda}, \boldsymbol{g}\in\tilde{\boldsymbol{\mathcal{G}}}} \widehat{R}_{[\textit{-}k]}\big(q,f,\boldsymbol{g};\widehat{{\boldsymbol{b}}}_{[\textit{-}k]}\big)\notag \\\widehat{f}_{[\textit{-}k]},\widehat{\boldsymbol{g}}_{[\textit{-}k]}&=\argmax_{f\in \tilde{\mathcal{F}}_{\lambda},\boldsymbol{g}\in\tilde{\boldsymbol{\mathcal{G}}}_{\lambda}} \min_{q\in\Delta^M}\widehat{R}_{[\textit{-}k]}\big(q,f,\boldsymbol{g};\widehat{{\boldsymbol{b}}}_{[\textit{-}k]}\big),
 \label{equ:gdro}
 \end{align} 
 where $\widehat{R}_{[\textit{-}k]}(q,f,\boldsymbol{g};\boldsymbol{b}):=\sum_{m=1}^Mq_m\widehat{\mathbb{E}}_{[\textit{-}k]}\supm\big[\ell\{Y,f(X,Z)+g\supm(Z)\}-\ell\{Y,b\supm(Z)\}\big]$.
 }
Construct the test statistic as $\widehat{I}_X=K^{-1}\sum_{k=1}^K\widehat{I}_{X,[k]}$ where  $\widehat{I}_{X,[k]}=\widehat{R}_{[k]}\big(\widehat{q}_{[\textit{-}k]},\widehat{f}_{[\textit{-}k]},\widehat{\boldsymbol{g}}_{[\textit{-}k]};\widehat{{\boldsymbol{b}}}_{[\textit{-}k]}\big)$, with its empirical standard error $\widehat{{\rm SE}}$ given by
\begin{equation}
\widehat{{\rm SE}}^2=\frac{1}{K}\sum_{k=1}^K \widehat{q}_{[\textit{-}k]}{\trans} \diag\Big\{\frac{1}{n_{m}}\widehat{\mathbb{V}}\supm_{[k]}\big(\ell\{Y,\widehat{f}_{[\textit{-}k]}(X,Z)+\widehat{g}\supm_{[\textit{-}k]}\} - \ell\{Y,\widehat{b}\supm_{[\textit{-}k]}\}\big)\Big\}_{m\in[M]} \widehat{q}_{[\textit{-}k]},
\label{equ:var}
\end{equation}
where $\widehat{\mathbb{V}}\supm_{[k]}(\cdot)$ represents the sample variance operator on $\bold{D}_{[k]}\supm$.

\Return{} $\widehat{I}_X$ and $(1-\alpha)\%$ confidence interval (CI): $\big[\widehat{I}_{X}-\Phi^{-1}(1-\alpha/2)\widehat{{\rm SE}},~\widehat{I}_{X}+\Phi^{-1}(1-\alpha/2)\widehat{{\rm SE}}\big]$.
\caption{\label{algo1} Outline of MIMAL.}
\end{algorithm}

Learners in Algorithm \ref{algo1} are fitted from the ML spaces $\tilde{\mathcal{F}}_{\lambda}$ and $\tilde{\boldsymbol{\mathcal{G}}}_{\lambda}$ with hyperparameters in $\lambda$. For example, $\tilde{\mathcal{F}}_{\lambda}=\{x\trans\beta:\|\beta\|_1\leq\lambda\}$ stands for lasso or $\tilde{\mathcal{F}}_{\lambda}$ can be a class of neural networks with some pre-specified architecture regularized with $\lambda$. More examples used in our numerical studies are given in Appendix. The ML method specified with $\tilde{\mathcal{F}}_{\lambda}$ and $\tilde{\boldsymbol{\mathcal{G}}}_{\lambda}$ are supposed to provide good estimation for the population models $\argmax_{f'\in\mathcal{F},\boldsymbol{g}'\in\boldsymbol{\mathcal{G}}} R(q,f',\boldsymbol{g}';\bar{\boldsymbol{b}})$ with an arbitrary source-mixture weight $q\in\Delta\supm$; see more details in Assumption \ref{minimum_rate}. Asymptotic unbiasedness and normality of $\widehat{I}_X$ are established in Theorem \ref{thm:normal:g} and equation (\ref{equ:var}) provides a standard moment estimator for the SE of $\widehat{I}_X$ based on this theorem.

\begin{remark}
The asymptotic normality of $\widehat{I}_X$ holds only when the true $I_X^*$ stays away from $0$. When $I_X^*=0$ or diminishes very fast to $0$, $\widehat{I}_X$ will be degenerated just like the chi-squared test statistic; see our simulated example in Figure \ref{fig:null}. This issue has frequently occurred in global inference literature like \cite{chakravarti2019gmm}, \cite{variance_inflation} and \cite{park2023robustui}. To ensure validity in the presence of such degeneration, we recommend a simple variance inflation strategy that uses an enlarged $\widehat{{\rm SE}}^2_{\tau}:=\widehat{{\rm SE}}^2+{\tau}/{\min_{m\in[M]}\{n_m\}}$ to replace $\widehat{{\rm SE}}^2$ for the interval estimate in Algorithm \ref{algo1}, where $\tau$ is a small and positive constant taken as $0.1$ or so. This is shown to attain good numerical performance in Section \ref{sec:sim:null}.
\label{rem:2}
\end{remark}

\begin{remark}
Our method can be naturally extended to handle the paired sampling design with $n_1=\ldots=n_M$ and each subject $i$ has $M$ dependent observations $\{(y\supm_i,x\supm_i,z\supm_i):m\in[M]\}$. The data used in our real example owns such a structure that observations on each date (subject) are collected at multiple locations. In this scenario, $\widehat{I}_X$ obtained using Algorithm \ref{algo1} preserves asymptotic normality with a different form of asymptotic variance. Correspondingly, for inference of $I^*_X$ in this situation, we only need to modify the empirical SE in (\ref{equ:var}) as 
\[
\widehat{{\rm SE}}^2=\frac{1}{n_1K}\sum_{k=1}^K \widehat{q}_{[\textit{-}k]}{\trans}\widehat{{\rm Cov}}_{[k]}\Big(\big[\ell\{Y,\widehat{f}_{[\textit{-}k]}(X,Z)+\widehat{g}\supm_{[\textit{-}k]}\} - \ell(Y,\widehat{b}\supm_{[\textit{-}k]})\big]_{m\in[M]}\Big)\widehat{q}_{[\textit{-}k]},
\]
where $\widehat{{\rm Cov}}_{[k]}$ represents the empirical covariance operator on the $k$-th fold. 
\label{rem:3}
\end{remark}

\subsection{Gradient-based optimization}\label{sec:method:grad}

We now introduce optimization procedures to extract the solution $\widehat{q}_{[\textit{-}k]},\widehat{f}_{[\textit{-}k]},\widehat{\boldsymbol{g}}_{[\textit{-}k]}$ of (\ref{equ:gdro}). A natural and commonly used generalization of gradient descent to such adversarial optimization problems is known as \textit{gradient-descent-ascent} (\texttt{GDA}). Suppose the ML technique parametrizes learners ${f}(x,z)={f}_{\theta_{f}}(x,z)$ and ${\boldsymbol{g}}(z) = {\boldsymbol{g}}_{\theta_{\boldsymbol{g}}}(z)$ by $(\theta_{f},\theta_{\boldsymbol{g}})$. Then at each iteration, \texttt{GDA} takes a gradient-descent in the parameter space $(\theta_{f},\theta_{\boldsymbol{g}})$ of the ML models $(\widehat{f}_{\theta_{f}},\widehat{\boldsymbol{g}}_{\theta_{\boldsymbol{g}}})$ to increase the prediction reward specified by the current adversarial weight $q$, followed by a gradient-ascent in $q$ with projection onto $\Delta^{M}$, to decrease the reward with the current $(\widehat{f}_{\theta_{f}},\widehat{\boldsymbol{g}}_{\theta_{\boldsymbol{g}}})$. 

Traditional GDA for adversarial machine learning suffers from limit cycling or even non-convergence without strict convexity and strict concavity guarantees. This significant optimization issue motivated the improvement of \texttt{GDA} with a \textit{two-timescale update rule} (\texttt{TTUR-GDA}) introduced for learning Actor-Critic methods \citep{ActorCritic-TTUR} and generative adversarial networks \citep{GAN-TTUR}. This \texttt{TTUR-GDA} algorithm is used in our framework to address the adversarial learning task (\ref{equ:gdro}).

Let $\theta_{f,t}$ and $\theta_{\boldsymbol{g},t}$ be the updated parameterizations of $\widehat{f}$ and $\widehat{\boldsymbol{g}}$ at iteration $t$.

\begin{algorithm}[H]
\caption{Two-Timescale GDA (\texttt{TTUR-GDA})}
\begin{algorithmic}
\REQUIRE initialization \( (q_0, \theta_{f,0},\theta_{\boldsymbol{g},0},\boldsymbol{b}) \), two step size series \( \eta_q(t) \) and  \(\eta_{f,\boldsymbol{g}}(t)\) with different scales, and the iteration length $T$.
\FOR{\( t = 1, 2, \ldots, T \)}
    \STATE Update \( (\theta_{f,t},\theta_{\boldsymbol{g},t}) \leftarrow (\theta_{f,t-1},\theta_{\boldsymbol{g},t-1}) - \eta_{f,\boldsymbol{g}}(t) \nabla_{\theta_f,\theta_{\boldsymbol{g}}}  \widehat{R}(q_{t-1}, \widehat{f}_{\theta_{f,t-1}},\widehat{\boldsymbol{g}}_{\theta_{\boldsymbol{g},t-1}};\widehat{{\boldsymbol{b}}}) \).
    \STATE Update \( q_t \leftarrow \mathcal{P}_{\Delta^M}\big\{q_{t-1} + \eta_q(t)\nabla_{q}\widehat{R}(q_{t-1}, \widehat{f}_{\theta_{f,t-1}},\widehat{\boldsymbol{g}}_{\theta_{\boldsymbol{g},t-1}};\widehat{{\boldsymbol{b}}})\big\} \), where $\mathcal{P}_{\Delta^M}(\cdot)$ represents the \texttt{projsplx} step introduced in Appendix. 
\ENDFOR
\RETURN  \((\widehat{q},\widehat{f},\widehat{\boldsymbol{g}})  = (q_T,\widehat{f}_{\theta_{f,T}},\widehat{\boldsymbol{g}}_{\theta_{\boldsymbol{g},T}})\).
\end{algorithmic}
\label{ttur-gda}
\end{algorithm}

In \texttt{TTUR-GDA}, $q$ is updated using the projected gradient-ascent algorithm described in Appendix with its step size $\eta_q(t)$ distinguished from $\eta_{f,\boldsymbol{g}}(t)$ used for the gradient update of ML models $(f_{\theta_f},\boldsymbol{g}_{\theta_{\boldsymbol{g}}})$. As suggested by \cite{GDA}, the step size $\eta_q(t)$ for the low-dimensional and convex weights should be chosen to dominate the step size $\eta_{f,\boldsymbol{g}}(t)$ for the complex and possibly non-concave ML estimators. In specific, when $\eta_{f,g}(t)$ is roughly on the cubic order of $\eta_q(t)$, \texttt{TTUR-GDA} ensures provable convergence to some (local) saddle-point solution \citep[Theorem 4.9]{GDA}. Intuitively, since the problem $\min_q R(q,f,\boldsymbol{g};\bar{\boldsymbol{b}})$ with fixed $f,\boldsymbol{g}$ is convex and easier to optimize, one could adopt a larger step size for the convex optimization of the optimal weight $q$ and make the GDA optimization process marginal in learning $(f,\boldsymbol{g})$ only. In our framework, \texttt{TTUR-GDA} allows the implementation of general ML methods with differentiable objective functions. This includes not only parametric regression and kernel methods, but also deep neural networks and gradient-boosted tree models. In numerical studies, we implement \texttt{TTUR-GDA} on \texttt{PyTorch} with the stochastic gradient optimizer. 

%For gradient-boosting, this can also be done by customizing loss functions in the libraries \texttt{XGBoost} and \texttt{LightGBM}. 

\section{Theoretical Analysis}\label{sec:thm}

\subsection{Setup without $Z$}\label{sec:thm:noZ}

Similar to Section \ref{sec:framework}, we start from the simplified problem without the baseline covariates $\{Z\supm\}_{m\in[M]}$ and its associated effect $\{g\supm\}_{m\in[M]}$. We will then show that the asymptotic analysis of the general case follows from it. Our population objective is now: 
\begin{align}
\label{simple_obj}
    {I}^*_X \coloneqq \max_{f\in\mathcal{F}}\min_{q\in\Delta^M}\left[\sum_{m=1}^M q_m\mathbb{E}\supm(\ell\{(Y,f(X)\}-\ell\{Y,0\}) \right].
\end{align} 
For simplicity, we drop the subscripts in $[k]$ and $[\textit{-}k]$ related to cross-fitting, e.g., the solution of the cross-fitted (\ref{equ:gdro}) is written as $(\widehat{q},\widehat{f})$ in this section. Suppose that the samples $\{y_i\supm,x_i\supm\}_{i=1,\ldots,n_m}$ are independent across the sources $m=1,\ldots,M$. Our framework also accommodates the paired sampling design with across-source-dependence as discussed in Remark \ref{rem:3}. Denote by $n:=n_1$ and $\rho_m \coloneqq {n_m}/{n_1} = n_m/n$. For simplicity of our presentation, we suppose that sample sizes of all sources grow in the same asymptotic order, that is, each of the finite sample size ratio $\rho_m$ converges to a limit $\bar{\rho}_m$ as $n\to\infty$ where $0<\bar{\rho}_m<\infty$.

Recall that the empirical version of $R(q,f)$ is defined as
\begin{align*}
    \widehat{R}(q,f)&\coloneqq \sum_{m=1}^M q_m \left(\dfrac{1}{n_m}\sum_{i=1}^{n_m}\ell\{y_i\supm,f(x_i\supm)\} - \ell\{y_i\supm,0\} \right),\\
    & = \sum_{m=1}^M q_m\left(\dfrac{\rho_m^{-1}}{n}\sum_{i=1}^{n_m} \ell\{y_i\supm,f(x_i\supm)\} - \ell\{y_i\supm,0\}\right).
\end{align*}

Let $V:\mathcal{F}\to \mathbb{R}$ be an arbitrary G\^ateaux differentiable function, and define the directional G\^ateaux derivative of $V$ at $f$ in the direction of $\phi\in \mathcal{F}$:
\begin{equation}
\label{gateaux_derivative}
    d(V(f);\phi)\coloneqq \lim_{\epsilon\to 0} \dfrac{V(f+\epsilon \phi) - V(f)}{\epsilon} = \dfrac{d}{d\epsilon} V(f+\epsilon \phi)\Bigr\rvert_{\epsilon=0}.
\end{equation}
Note that for each $f\in\mathcal{F}$ the mapping $d(V(f);\,\cdot\,)$ is also a real-valued function defined on the space $\mathcal{F}$. Higher order derivative are defined as $d^{k}(V(f);\phi) \coloneqq \dfrac{d^{k}}{d\epsilon^{k}} V(f+\epsilon \phi)\Bigr\rvert_{\epsilon=0}$.

We make the following regularity conditions to facilitate our theoretical analysis.
\begin{assumption}
Covariates $X\supm$ from each $m\in[M]$ is supported on a common set $\mathcal{X}\subset \mathbb{R}^p$. There exists some constant $C>0$ such that for every measurable set $A\subset \mathcal{X}$, $C^{-1}<{\mathbb{P}_X^{(m)}}(A)/{\mathbb{P}_X^{(l)}}(A)<C$ for any two $m,l\in[M]$, where $\mathbb{P}_X^{(m)}(A)$ represents the probability measure of $A$ on source $m$.
\label{asu:1}
\end{assumption}
\begin{assumption}[Strict convexity]
\label{restriced_convexity}
    The function $\lambda^*(q) \coloneqq \max_{f\in\mathcal{F}}R(q,f)$ is strictly convex on the simplex $\Delta^M$.
\end{assumption}

We say that a function $F :\mathcal{F} \to\mathbb{R}$ is concave if for any $f_1,f_2\in\mathcal{F}$ and any $\alpha \in [0,1]$, we have $F(\alpha f_1+(1-\alpha)f_2) \geq \alpha F(f_1) + (1-\alpha)F(f_2)$. It is strictly concave if the inequality is strict whenever $f_1\neq f_2$. 
\begin{assumption}[Regularity of $\ell$]
\label{regularity}
    For each $m\in[M]$, the second moment of $\ell\{Y,f(X)\}$ is finite for all $f$ in a neighborhood of $\bar{f}$ and the function $f \mapsto \mathbb{E}\supm \ell\{Y,f(X)\}$ is continuous and strictly concave in $f\in \mathcal{F}$. The continuous function $(q,f) \mapsto R(q,f)$ is twice G\^{a}teaux differentiable at the saddle point $(\bar{q},\bar{f})$, specifically both $d(R(\bar{q},\bar{f});(q',f'))$ and $d^2(R(\bar{q},\bar{f});(q',f'))$ as defined in (\ref{gateaux_derivative}) exist and are finite for all $(q',f') \in \Delta^M\times \mathcal{F}$. 
\end{assumption}

\begin{assumption}
\label{asm:1A}
For every source $m\in[M]$, one can interchange the first and second order G\^ateaux differentiation in $f$ and expectation, e.g. for first order differentiation $d(\mathbb{E}\supm({\ell}\{{Y},f({X})\});\phi) = \mathbb{E}\supm d({\ell}\{{Y},f({X})\};\phi)$ for $f,\phi\in\mathcal{F}$. 
\end{assumption}

Assumption \ref{asu:1} means that the covariate distributions overlap across the sources. Assumptions \ref{restriced_convexity} and \ref{regularity} imply that the saddle point $(\bar{q},\bar{f})$ is unique. This is because strict convexity of $\lambda^*(q)$ implies uniqueness of $\bar{q}$, which in turn implies uniqueness of $\argmax_f R(\bar{q},f)$ by Assumption \ref{regularity}. In addition, Assumption \ref{asm:1A} concerns the interchangeability of G\'{a}teaux differentiation and expectation of the population objective. Note that both Assumption \ref{regularity} and Assumption \ref{asm:1A} are standard and hold in general cases such as the exponential family log-likelihood. 

For each given $q\in\Delta^M$, we denote $\bar{f}_{\text{ERM}}(x;q):=\argmax_{f\in \mathcal{F}} R(q,f)$ as the population version of the empirical reward maximizer (ERM) on the mixture of the sources with the weights $q$. Correspondingly, we define $\widehat{f}_{\text{ERM}}(x;q)=\max_{f\in \tilde{\mathcal{F}}_{\lambda}} \widehat{R}(q,f)$ as its estimator extracted on the $q$-mixed source samples from the ML space $\tilde{\mathcal{F}}_{\lambda}$. 

\begin{assumption}[Lipschitz implicit function]
\label{lipschitz}
The ERM $\bar{f}_{\text{ERM}}(x;q)$ is Lipschitz continuous in $q$, that is, $\mathbb{E}\supm[\bar{f}_{\text{ERM}}(X;q_1) - \bar{f}_{\text{ERM}}(X;q_2)]^2\leq L\lVert q_1-q_2\rVert_2^2$, for some constant $L>0$ and all $m\in[M]$.
\end{assumption}

In Proposition \ref{prop:asu:4}, we justify that Assumption \ref{lipschitz} holds for the least square and logistic regressions. In general, this can be extended to other cases with smooth and regular $\ell$.

\begin{prop}
\label{prop:asu:4}
    When $\ell(y,u)=-{(y-u)^2}$, the ERM $\bar{f}_{\text{ERM}}(x;q)$ is Lipschitz continuous in $q$. When $\ell(y,u) = yu-\log{(1+e^u)}$, $\bar{f}_{\text{ERM}}(x;q)$ is Lipschitz in $q$ if $\bar{f}_{\text{ERM}}(x;q)$ is bounded in values.
\end{prop}

\begin{assumption}[Convergence rate of ML]
\label{minimum_rate}
The source-mixture ERM satisfies that 
\[
\sup_{q\in \Delta^M,m\in[M]}\Big(\mathbb{E}\supm_X [\bar{f}_{\text{ERM}}(X;q)-\widehat{f}_{\text{ERM}}(X;q)]^2\Big)^{1/2}= o_p(n^{-1/4}),
\]
where $\mathbb{E}\supm_X$ denotes the expectation operator with respect to a realization of $X$ on the source $m$ independent of $\widehat{f}_{\text{ERM}}$.
\end{assumption}

Assumption \ref{minimum_rate} requires the machine learner to achieve $o_p(n^{-1/4})$-convergence in terms of the $\ell_2$-error on the ERM task on each $q$-weighted mixture of the sources. The same convergence rate is required by recent ML-based inference methods for the single-source scenario \citep[e.g.]{doubleML,general-framework}. Note that for least squares regression with $\ell(y,u)=-{(y-u)^2}$, we have $\bar{f}_{\text{ERM}}(x;q)=\sum_{m=1}^M q_m \bar{f}_{\text{ERM},m}(x)$, where $\bar{f}_{\text{ERM},m}(x)$ represents the ERM on source $m$. In this case, Assumption \ref{minimum_rate} holds whenever the ML method attains $o_p(n^{-1/4})$-convergence on the ERM task of every source. For general $\ell(y,u)$, $o_p(n^{-1/4})$-convergence of the ML on every source does not naturally imply Assumption \ref{minimum_rate}. Nevertheless, structural complexity properties like sparsity and smoothness on mixture of the sources tend be inheritable from those on each single source. Thus, one could expect ML approaches to perform similarly well on learning $\bar{f}_{\text{ERM}}(q)$ and $\bar{f}_{\text{ERM},m}$. 

%Suppose the probability density function of $\mathbb{P}\supm_{Y|X}\times\mathbb{P}\supm_{X}$ has an $r$-th derivative for all $m\in[M]$. Then the density function of $\sum_{m=1}^Mq_m\mathbb{P}\supm_{Y|X}\times\mathbb{P}\supm_{X}$ also has an $r$-th derivative for all $q\in\Delta^M$.

More sophisticated than ML-agnostic inference for single-source variable importance \citep{general-framework}, justification for the $n^{-1/2}$-consistency and asymptotic normality of $\widehat{I}_X$ relies on a key extension of the Neyman orthogonality from the ERM problem to the adversarial learning task as presented in Lemma \ref{lem:1}.

\begin{lemma}[Neyman orthogonality in group adversarial learning]
\label{lem:1}
    Suppose Assumption \ref{regularity} and \ref{asm:1A} hold. Let $(\bar{q},\bar{f})$ be the saddle point solution to the objective $\max_f \min_q R(q,f)$ as defined in (\ref{equ:nash:pop}) (without $Z$). For any $(q',f')\in\Delta^M\times \mathcal{F}$ such that $q'$ has the same support (i.e., the same set of non-zero elements) as $\bar{q}$, the first order G\^{a}teaux expansion of $R(q,f)$ at $(\bar{q},\bar{f})$ in the direction of $(\bar{q}-q',\bar{f}-f')$ is zero. Specifically,
    \[
    d(R(\bar{q},\bar{f});(q'-\bar{q},f'-\bar{f}))=\dfrac{d}{d\epsilon} R(\bar{q}+\epsilon (q'-\bar{q}), \bar{f}+\epsilon (f'-\bar{f}))\Bigr\rvert_{\epsilon = 0} = 0. 
    \]

\end{lemma}

\begin{remark}
In Lemma \ref{lem:1}, we consider sufficient condition under which the G\^{a}teaux derivative of the objective $R(q,f)$ vanishes at $(\bar{q},\bar{f})$ in the direction of approximation residuals. In particular, we require that the support (nonzero components) of $q'$ to coincide with the support of $\bar{q}$, which seems to be restrictive. However, when $q'=\widehat{q}$, it is guaranteed that this sufficient condition is met for $q'$ when $n$ is large enough, except for an edge case discussed in details in the proof of Theorem \ref{main_theorem}. 
\end{remark}

Importantly, we do not directly impose convergence assumptions on the adversarial learning task (\ref{equ:gdro}), considering that it has not been studied as broadly as the ERM problem for various ML approaches. Instead, we leverage Assumption \ref{minimum_rate} to justify the joint $o_p(n^{-1/4})$-convergence of the empirical saddle point $(\widehat{q},\widehat{f})= \max_{f\in\tilde{\mathcal{F}}_{\lambda}}\min_{q\in\Delta^M} \widehat{R}(q,f)$ in Lemma \ref{lem:2}. As a benefit of this, one could further justify our Assumption \ref{minimum_rate} through comprehensive literature like \cite{negahban2012unified} for Lasso, \cite{aronszajn1950theory} for kernel methods, and \cite{schmidt2020nonparametric,farrell2021deep} for deep neural networks.

\begin{lemma}[Convergence of the adversarial learner.]
\label{lem:2}
    If Assumptions \ref{asu:1} -- \ref{minimum_rate} hold, then the saddle point satisfies that $\big(\mathbb{E}_X\supm[\widehat{f}(X) - \bar{f}(X)]^2\big)^{1/2}\lesssim o_p(n^{-1/4})$ and $\lVert \widehat{q} - \bar{q}\rVert_2 \lesssim o_p(n^{-1/4})$.
\end{lemma}

In the following, we briefly introduce how Lemmas \ref{lem:1} and \ref{lem:2} can be used to address the first and second order errors in $\widehat{I}_X$ respectively. Heuristically, we can expand $\widehat{I}_X$ with respect to $\widehat{q}$ and $\widehat{f}$ to derive that 
\[
\widehat{I}_X=\widehat{R}(\widehat{q},\widehat{f})\approx\widehat{R}(\bar{q},\bar{f})+d(\widehat{R}(\bar{q},\bar{f});(\widehat{q}-\bar{q},\widehat{f}-\bar{f}))+\mbox{Sec}(\widehat{q}-\bar{q},\widehat{f}-\bar{f}),
\]
where $\mbox{Sec}(\widehat{q}-\bar{q},\widehat{f}-\bar{f})$ denotes the second order error of $\widehat{q}$ and $\widehat{f}$. Then by the Neyman orthogonality in Lemma \ref{lem:1}, for any $(\delta,\phi)\in \mathbb{R}^M\times\mathcal{F}$, $d(\widehat{R}(\bar{q},\bar{f});(\delta,\phi))$ has a zero mean and, thus, concentrating to zero in the parametric rate. Combining this with the consistency of $(\widehat{q},\widehat{f})$, we can remove the first order bias term $d(\widehat{R}(\bar{q},\bar{f});(\widehat{q}-\bar{q},\widehat{f}-\bar{f}))$ compared to the leading term $\widehat{R}(\bar{q},\bar{f})$. This is an automatic procedure without requiring any extra bias-correction steps. Also, the $o_p(n^{-1/4})$-convergence of $(\widehat{q},\widehat{f})$ established in Lemma \ref{lem:2} ensures the second order error $\mbox{Sec}(\widehat{q}-\bar{q},\widehat{f}-\bar{f})= \{o_p(n^{-1/4})\}^2=o_p(n^{-1/2})$, which is negligible compared to $\widehat{R}(\bar{q},\bar{f})$. Based on these, we justify the asymptotic properties of $\widehat{I}_X$ in Theorem \ref{main_theorem}.

\begin{theorem}
\label{main_theorem}
    If Assumptions \ref{asu:1} -- \ref{minimum_rate} hold with $I_X^*>0$, and further assuming the samples are independent across the sources, then the fitted MIMAL variable importance $\widehat{I}_{X}$ is consistent with $I_X^* = R(\bar{q},\bar{f})$, and satisfies that
    \[
    \sqrt{n}\big(\widehat{I}_{X} - I_X^*\big) \leadsto N(0,\sigma^2),\quad \mbox{where}\quad\sigma^2=\bar{q}\trans \diag{\{\bar{\sigma}^2_{(m)}/\bar{\rho}_m\}}\bar{q},
    \]
    where $\bar{\sigma}^2_{(m)} \coloneqq \var[\ell \{ Y\supm, \bar{f}(X\supm)\} - \ell\{Y\supm,0\}]$ and $\leadsto$ stands for weak convergence. When the objective value $I_X^* = 0$, we have $\widehat{I}_X-I_X^* = o_p(n^{-1/2})$. 
\end{theorem}

\begin{corollary}[Interval Estimation]
\label{interval_estimate}

Suppose all assumptions in Theorem \ref{main_theorem} hold.

\begin{enumerate}
    \item If $I_X^* > 0$, the empirical variance $n\widehat{{\rm SE}}^2$ defined in (\ref{equ:var}) is a consistent estimator of $\sigma^2$ and the $(1-\alpha)$-confidence interval $
\ci(\alpha) \coloneqq  \big[\widehat{I}_{X} \pm \Phi^{-1}(1-\alpha/2) \widehat{{\rm SE}} \big]
$ satisfies that, \[
\lim_{n\to\infty}\mathbb{P}\big({I}_{X}^* \in \ci(\alpha) \big) = (1-\alpha).
\]
    \item If instead we use the variance inflation strategy defined in Remark \ref{rem:2} and the inflated CI $\ci_\tau(\alpha) \coloneqq \big[\widehat{I}_X \pm\Phi^{-1}(1-\alpha/2)\widehat{\rm SE}_\tau\big]$, then it satisfies for any $I_X^*\geq 0$ that, 
    \[
\lim_{n\to\infty} \mathbb{P}\big(I_X^* \in \ci_\tau(\alpha)\big) \geq 1-\alpha. {\color{red} }
\]
\end{enumerate}
\end{corollary}

\begin{remark}
The asymptotic normality of $\widehat{I}_{X}$ lies on the uniqueness of $(\bar{q},\bar{f})$. This is guaranteed if the convexity of the function $\lambda^*(q)$ is strict as imposed in Assumption \ref{restriced_convexity}. In some subtle situations such as when two sources having an identical data distribution, i.e., $\mathbb{P}^{(m_1)}_{Y|X}\times\mathbb{P}^{(m_1)}_X \overset{d}{=} \mathbb{P}^{(m_2)}_{Y|X}\times \mathbb{P}^{(m_2)}_X$, uniqueness of $\bar{q}$ fails to hold as $q_{m_1}$ and $q_{m_2}$ satisfied a linear dependency $q_{m_1}+q_{m_2}=c$, for some $0<c<1$, even though $I_X^*$ and $\bar{f}$ are still uniquely identifiable. This implies there can be uncountably many different solutions in $\bar{q}$.  When Assumption \ref{restriced_convexity} fails there is no limiting variance of the $\widehat{I}_X$ as it depends on $q$. Adding a ridge penalty term on $q$ is proposed by \cite{Zijian_maxmininfer} to address this issue. In this way, the population reward function becomes 
\begin{equation}
R_{\delta}(q,f) \coloneqq \sum_{m=1}^M q_m\mathbb{E}\supm(\ell\{Y,f(X)\} - \ell\{Y,0\}) + \delta\lVert q \rVert_2^2,
\label{equ:ridge}
\end{equation}
whose saddle-point solution is unique. In our real-world study in Section \ref{real-data}, we take $\delta=0.001$ to ensure training stability. 

\label{rem:non-uni}
\end{remark}

In addition, we demonstrate that cross-fitting is no longer necessary when the model for $f$ belongs to some Donsker class over a compact normed model space $\tilde{\mathcal{F}}_{\lambda}$, with specific examples including parametric model and kernel ridge regression introduced in Appendix. In this scenario, one could justify Theorem \ref{main_theorem} by generalizing the tangential Hadamard directional differentiability \citep[Theorem 7.24]{shapiro} of the maximin function $\psi(f) = \max_f\min_q R(q,f)$ to normed vector spaces, and using the argument of \citep[Theorem 5.10]{shapiro} that the in-sample maximin objective value is asymptotically normal. See the following corollary for more details. 

\begin{corollary}
    The statement of Theorem \ref{main_theorem} applies to the inference of MIMAL for both finite-dimensional parametric models and kernel ridge regression introduced in Appendix, without requiring cross-fitting.
\end{corollary}

\subsection{Extension to setup with $Z$}\label{sec:thm:withZ}

Now we consider the general scenario with baseline covariates $Z$ and the population objective function is as defined in (\ref{equ:pop:maxmin}). As discussed in Remark \ref{rem:1}, this form raises the possibility of non-identifiability in $(f,\boldsymbol{g})$. In the following, we shall explain that this is not fatal as $f+\boldsymbol{g}$ and $I^*_X$ are generally identifiable. Denote by $\boldsymbol{h} \coloneqq (h^{(1)},\ldots,h^{(M)})\in \mathcal{H}^{(1)}\times\ldots \times\mathcal{H}^{(M)} \eqqcolon \boldsymbol{\mathcal{H}}$ where $\mathcal{H}\supm \coloneqq \{f+g\supm \mid f\in\mathcal{F},g\supm\in\mathcal{G}\supm\}$. Let ${\boldsymbol{\mathcal{H}}}^\dagger \coloneqq \{\boldsymbol{h}=(h^{(1)},\ldots,h^{(M)})\in\boldsymbol{\mathcal{H}}\mid \boldsymbol{h}= f + \boldsymbol{g} \text{ for }f\in\mathcal{F},~\boldsymbol{g}\in\boldsymbol{\mathcal{G}}\}$. The space ${\boldsymbol{\mathcal{H}}}^{\dagger}$ can be viewed as a structural subset of $\boldsymbol{\mathcal{H}}$ with its functions sharing the model $f$ to characterize the stable effect of $X$ and separately using $g\supm$ in each source $m$ for confounding adjustment. Then the population reward defined in (\ref{equ:pop:maxmin}) can be written as:
\[
R(q,\boldsymbol{h};\bar{\boldsymbol{b}})=R(q,\boldsymbol{h})=\sum_{m=1}^M q_m\big[\mathbb{E}\supm\ell\{Y,h\supm(X,Z)\}-\mathbb{E}\supm\ell\{Y,\bar{b}\supm(Z)\}\big].
\]
In Lemma \ref{lem:3}, we justify that the maximin problem constructed with $R(q,\boldsymbol{h};\bar{\boldsymbol{b}})$ has a unique solution $(\bar{q},\bar{\boldsymbol{h}})$ under the strict concavity of $R(q,\boldsymbol{h};\bar{\boldsymbol{b}})$ on the structural function space ${\boldsymbol{\mathcal{H}}}^{\dagger}$.
\begin{lemma}
\label{lem:3}
Under Assumption \ref{A1a} -- \ref{A2a} to be introduced below, $\max_{\boldsymbol{h}\in \boldsymbol{\mathcal{H}}^\dagger}\min_{q\in\Delta^M} R(q,\boldsymbol{h};\bar{\boldsymbol{b}})$ has a non-negative and finite optimal value and unique saddle point $(\bar{q},\bar{\boldsymbol{h}})$.
\end{lemma}

We list the required assumptions below. Let $\mathcal{F}$ and each $\mathcal{G}\supm$ be convex, closed and bounded. Assumptions \ref{A1a} -- \ref{A6a} are natural extensions of Assumptions \ref{asu:1} -- \ref{minimum_rate} to the scenario with confounding adjustment. Assumption \ref{A7a} is about the ML estimation convergence rate on each baseline model $b\supm$. 

\begin{assumptionA}{1}
Covariates $(X\supm,Z\supm)$ from each $m\in[M]$ is supported on a common set $\mathcal{X}\times\mathcal{Z}\subset \mathbb{R}^p\times \mathbb{R}^k$. There exists constant $C>0$ such that for every measurable set $A\subset \mathcal{X}\times \mathcal{Z}$, we have $C^{-1}<{\mathbb{P}_{X,Z}^{(k)}}(A)/{\mathbb{P}_{X,Z}^{(l)}}(A)<C$ for any two $l,k\in[M]$.
\label{A1a}
\end{assumptionA}

\begin{assumptionA}{2}
\label{A2a}
The function $\lambda^*(q) \coloneqq \max_{\boldsymbol{h}\in \boldsymbol{\mathcal{H}}^\dagger} R(q,\boldsymbol{h};\bar{\boldsymbol{b}})$ is strictly convex on $\Delta^M$.
\end{assumptionA}

\begin{assumptionA}{3}
For each $m\in[M]$, the second moment of $\ell\{Y,h\supm(X)\}$ is finite for all $h\supm$ in a neighborhood of $\bar{h}\supm$ and the function $h\supm \mapsto \mathbb{E}\supm \ell\{Y,h\supm(X)\}$ is continuous and strictly concave in $h\supm\in \mathcal{H}\supm$.The function $(q,\boldsymbol{h})\mapsto R(q,\boldsymbol{h};\bar{\boldsymbol{b}})$ is twice G\^{a}teaux differentiable at the unique saddle point $(\bar{q},\bar{\boldsymbol{h}})\in\Delta^M\times \boldsymbol{\mathcal{H}}^\dagger$ in all directions of $\Delta^M\times \boldsymbol{\mathcal{H}}$. 
\end{assumptionA}

\begin{assumptionA}{4}
\label{asm:2A}
For each of the site objective value $\mathbb{E}\supm(\ell\{Y,h\supm(X)\} - \ell\{Y,b\supm(X)\})$ in $h\supm$, it is satisfied that we may interchange first and second order G\^ateaux differentiation in $h\supm \in\mathcal{H}\supm$ and expectation. 
\end{assumptionA}

\begin{assumptionA}{5}
\label{A4a}
The argmax function $\bar{\boldsymbol{h}}_{\text{ERM}}(q)\coloneqq \argmax_{\boldsymbol{h}\in\boldsymbol{\mathcal{H}}^\dagger} R(q,\boldsymbol{h};\bar{\boldsymbol{b}})$ is Lipschitz continuous in $q\in\Delta^M$.
\end{assumptionA}
\begin{assumptionA}{6}
\label{A6a}
The source-mixture ERM estimators satisfy that 
\[
\sup_{q\in \Delta^M,m\in[M]}\Big(\mathbb{E}_X\supm [\bar{h}_{\text{ERM}}\supm(X;q)-\widehat{h}_{\text{ERM}}\supm(X;q)]^2\Big)^{1/2}= o_p(n^{-1/4}).
\]
\end{assumptionA}
\begin{assumptionA}{7}[Baseline Models]
\label{A7a}
It is satisfied that $\mathbb{E}_X\supm[\widehat{b}\supm(Z)-\bar{b}\supm(Z)]^2 \lesssim o_p(n_m^{-1/2})$ for each $m\in[M]$.
\end{assumptionA}

In the following Theorem \ref{thm:normal:g}, we extend the asymptotic properties of $\widehat{I}_X$ established in Theorem \ref{main_theorem} to the general setup with confounding adjustments.

\begin{theorem}
\label{thm:normal:g}
Suppose that Assumptions \ref{A1a} -- \ref{A7a} hold and $I_X^*>0$ with the samples being independent across the sources. The fitted MIMAL variable importance $\widehat{I}_{X}$ (with adjustment on $Z$) is consistent with $I^*_X=R(\bar{q},\bar{\boldsymbol{h}};\bar{\boldsymbol{b}}) = \max_{\boldsymbol{h}\in\boldsymbol{\mathcal{H}}^\dagger}\min_{q\in\Delta^M} R(q,\boldsymbol{h};\bar{\boldsymbol{b}})$ and satisfies that 
\[
\sqrt{n}\left(\widehat{I}_X - I^*_X\right) \leadsto N(0,\sigma^2),\quad \mbox{where}\quad \sigma^2= \bar{q}\trans \diag{\{\bar{\sigma}^2_{(m)}/\bar{\rho}_m\}}\bar{q},\]
where $\bar{\sigma}^2_{(m)} \coloneqq \var[\ell \{ Y\supm, \bar{h}\supm(X\supm,Z\supm)\} - \ell\{Y\supm,\bar{b}(Z\supm)\}]$. When $I_X^* = 0$, we have $\widehat{I}_X-I_X^* = o_p(n^{-1/2})$.
\end{theorem}

\section{Simulation Studies}\label{sec:sim}
\subsection{Overview}
In this section, we conduct comprehensive simulation studies to investigate the asymptotic normality of $\widehat{I}_X$ and the coverage probability (CP) of the $95\%$ CI produced by the MIMAL Algorithm \ref{algo1}, with various setups in the data generation mechanism and ML constructions. An overview of the simulation setups and results is given in Table \ref{tab:overview}. For all the numerical studies, the \texttt{TTUR-GDA} algorithm is implemented through the automatic differentiation package \texttt{PyTorch} in \texttt{Python}. Additional construction details of all our used ML models can be found in Appendix. We replicate $1000$ times on each simulation setting to estimate the CP of our method. 

\begin{table}[H]
    \centering
    \begin{tabular}{lllllll}
        \toprule
        Reward & Learner (\# of Params) & Nuisance & Null-model & Coverage \\
        \midrule
        $\ell_2$ & Lasso (50) & No  & No & 94.6\%\\
        $\ell_2$ & KRR (Gaussian) (1800) & Yes & No & 94\% \\
        Logistic & Neural Nets (57) & No & No & 94\% \\
        $\ell_2$ & GLM (6) & Yes  & Yes & 95.9\%\\
        Logistic & GLM (5) & Yes  & No & 95.8\% \\
        Poisson & Cubic B-Splines (24) & No  & No & 95.2\%\\
        \bottomrule
    \end{tabular}
    \caption{\label{tab:overview} A summary of simulation results. The reward functions are those based on likelihoods of Gaussian ($\ell^2$), binomial (Logistic) and Poisson distribution. `Nuisance' indicates whether baseline covariates $Z\supm$'s are included or not. `Null-model' indicates whether the population truth $I^*_X$ is $0$. `Coverage' is the coverage probability of our $95\%$ CI over $1000$ simulations. Experiment 5 and 6 are deferred to Appendix.}
\end{table}

\subsection{Simulation 1: Lasso Regression}

We generate $M=3$ data sources with $n_1=n_2=n_3=800$, $Y\supm = X\trans\theta\supm + \mathcal{N}(0,1)$, $\theta\supm\in\mathbb{R}^{50}$ and $X$ is $50$-dimensional uniformly distributed with independent components in the interval $[-3,3]$, i.e. $X\sim \mathcal{U}[-3,3]^{50}$. The $\theta\supm$ are designed to have $5$ significant nonzero components, and $45$ zero components, so that $\theta\supm = [\theta\supm_1,\theta\supm_2,\ldots,\theta\supm_5] + 45*[0]$, where $+$ here represents concatenating the two vectors and $*$ means concatenating the same vector for multiple times. In specific, we set
\begin{align*}
    \theta^{(1)} &= [5.78, -4.45, 1.26, 1.58, -1.14] + 45*[0];\\
    \theta^{(2)} &= [2.26, -1.05, 5.78, 6.43, -1.26] + 45*[0];\\
    \theta^{(3)} &= [1.83, -2.35, 1.34, 2.59, -6.45] + 45*[0].
\end{align*}
Without $Z\supm$, the baseline model is a null model only including the intercept term. The MIMAL objective reward function is set as $\ell(y,u)=-(y-u)^2$. Solution to the population maximin problem is $\bar{q} = [0.43,0.16,0.41]$ and $\bar{\theta} =  [3.60, -3.04, 2.03, 2.78, -3.32] + 45*[0]$. To estimate the model $f$, we use Lasso regression with the penalty coefficient set as $1/n_1$.

\begin{figure}[htbp]
    \centering
    \includegraphics[width=0.65\textwidth]{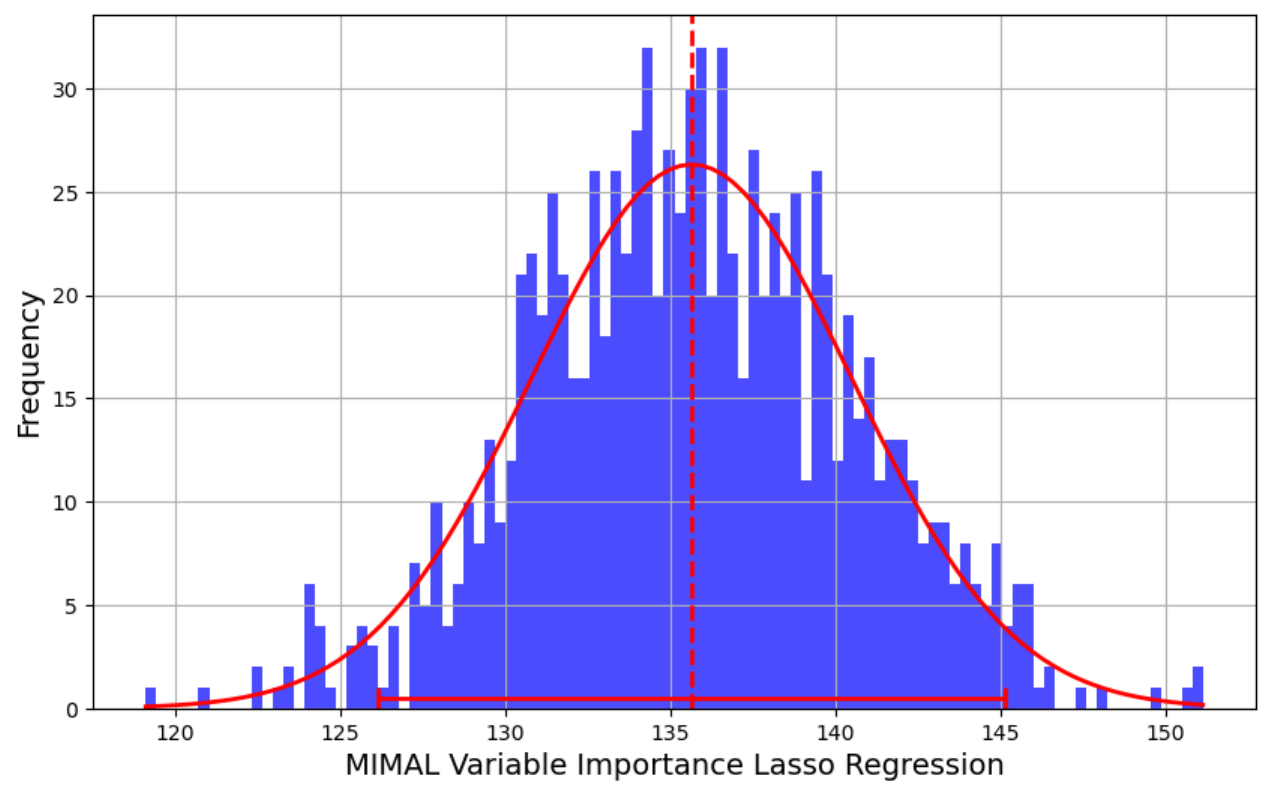}
    \caption{Empirical distribution of 1000 simulated MIMAL values in Simulation 1. The dotted vertical line indicates the true value. The horizontal red line indicates a $95\%$-interval around the truth. }
    \label{fig:lasso_density}
\end{figure}

As demonstrated in Figure \ref{fig:lasso_density}, our estimate of $\widehat{I}_X$ shows good normality with small bias. We have $94.6\%$ of the CI estimates to cover the true reward $135.243$. In addition, we find that the mean of $\widehat{q}$ is $[0.4321, 0.1609, 0.4071]$ and the mean of $\widehat{\theta}_{[1:5]}$ in our $\widehat{f}$ is $[3.601, -3.042,2.025, 2.782,-3.310]$. Both are close to the population saddle point. 

\subsection{Simulation 2: Kernel Ridge Regression}
\label{simul2}

The second simulation study implements KRR as described in Appendix. We include baseline covariates $\{Z_m\}_{m\in[M]}$ and again consider continuous $Y$ and $M=3$ sources. For each $m$, we generate $n_m=n=600$ samples of $Y\supm = X\trans\theta\supm + Z\trans\gamma\supm + \mathcal{N}(0,0.25)$, with $X \sim \mathcal{U}[-3,3]^3$, $Z \sim \mathcal{U}[-3,3]^2$, $[\theta^{(1)},\gamma^{(1)}] = [0.9,0.3,0.3] + [0.4,0.3]$, $[\theta^{(2)},\gamma^{(2)}] = [0.3,0.9,0.3] + [-0.3,0.2]$, and $[\theta^{(3)},\gamma^{(3)}] = [0.3,0.3,0.9] + [0.0,0.0]$. The true saddle-point solution is thus $\bar{\theta} = [0.5,0.5,0.5]$ with $\bar{\gamma}^{(m)} = \gamma\supm$ for $m=1,2,3$, whose MIMAL objective is approximated to $2.238$. 

The MIMAL objective is the usual $\ell_2$ reward function. We use the Gaussian (RBF) kernel $K(x_i,x_j|\sigma) = \exp{(-\sigma\lVert x_i-x_j\rVert_2^2)}$, where the tuning parameter $\sigma$ controls the `long-range' dependency between the feature maps. Large value of $\sigma$ localizes the prediction, and hence encourage training data interpolation and over-fitting. We study the asymptotic normality of our method with $\sigma = 0.1$ as well as the bias incursion when ranging $\sigma$ from $0.1$ to $0.5$. The regularization coefficient of KRR is set as $1/(10n)$. Similar to \cite{general-framework}, we also include both cross-fitted and non-cross-fitted versions of MIMAL for comparison.

\begin{figure}[htbp]
    \centering
    \subfigure{
        \includegraphics[width=0.479\textwidth]{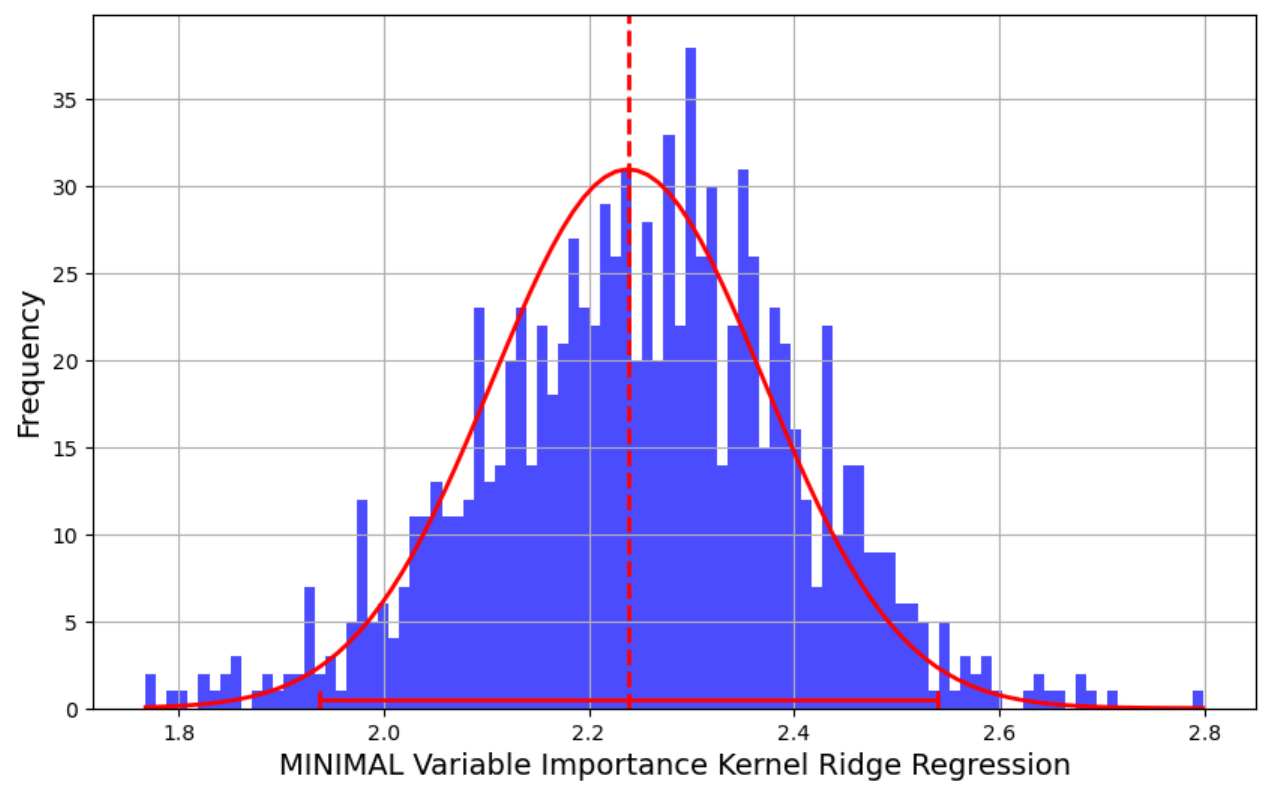}
    }
    \hfill
    \subfigure{
        \includegraphics[width=0.479\textwidth]{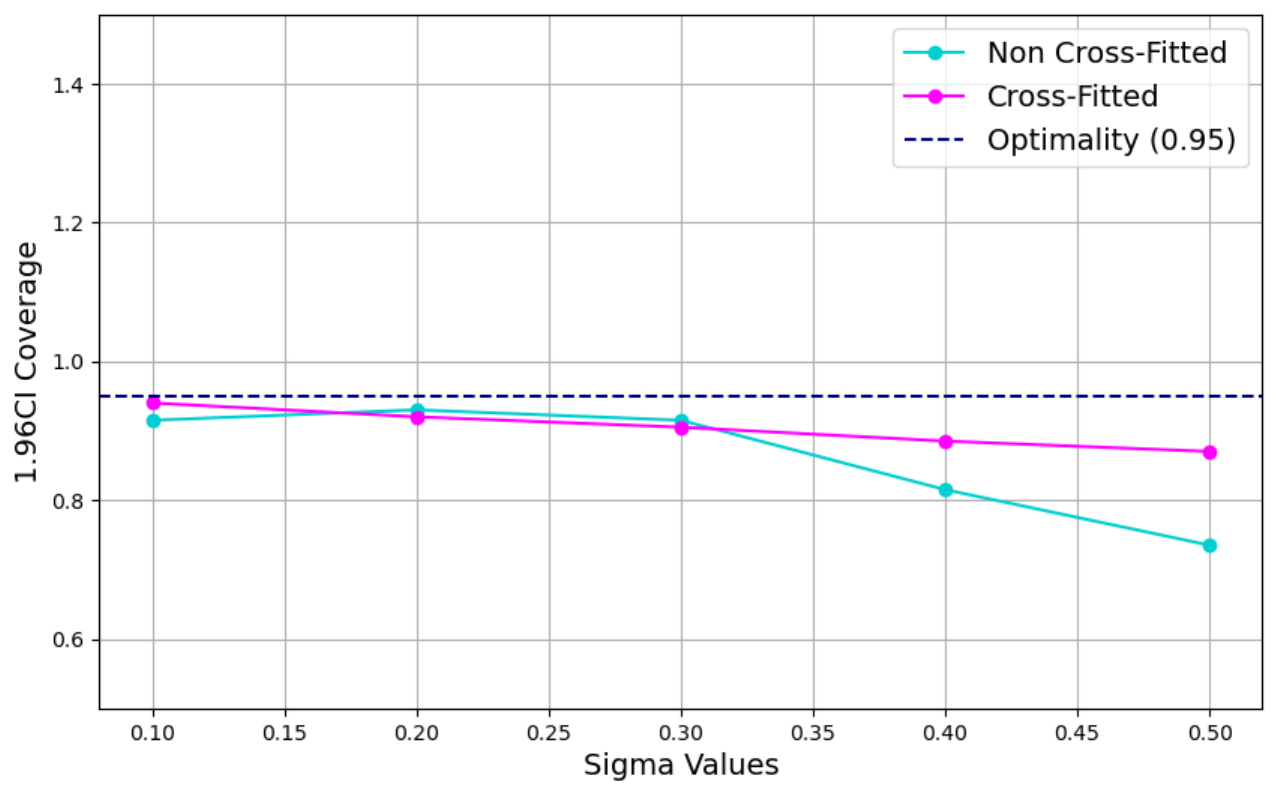}
    }
    \caption{Left figure: Empirical distribution of 1000 simulated MIMAL values in Simulation 2 with cross-fitted KRR ($\sigma=0.1$). The dotted vertical line indicates the true value. The horizontal red line indicates a $95\%$-interval around the truth. Right figure: Coverage probability with varying values of the tuning parameter $\sigma$ in cross-fitted and non-cross-fitted versions of MIMAL.}
    \label{fig:krr_2}
\end{figure}

As seen from the left one of Figure \ref{fig:krr_2}, the empirical distribution of $\widehat{I}_X$ is normal with a mean $2.250$ slightly shifted from the truth $2.238$. Nonetheless, the CP of MIMAL with KRR ($\sigma=0.1$ and cross-fitted) is still 94\% and close to the nominal level. From the right one in Figure \ref{fig:krr_2}, we observe that increasing $\sigma$ to $0.3$ or larger values could cause bias and under-coverage of our method. Also, the cross-fitted version of MIMAL attains significantly better coverage performance than its non-cross-fitted counterpart. For example, when $\sigma=0.4$, cross-fitted MIMAL attains a 90\% CP while the non-cross-fitted one only has it around 80\%. This demonstrates the effectiveness of cross-fitting in reducing over-fitting bias when using high-complexity ML models in our framework.

\subsection{Simulation 3: Neural Network Classification}

\begin{figure}[htbp]
    \centering
    \includegraphics[width=0.6\linewidth]{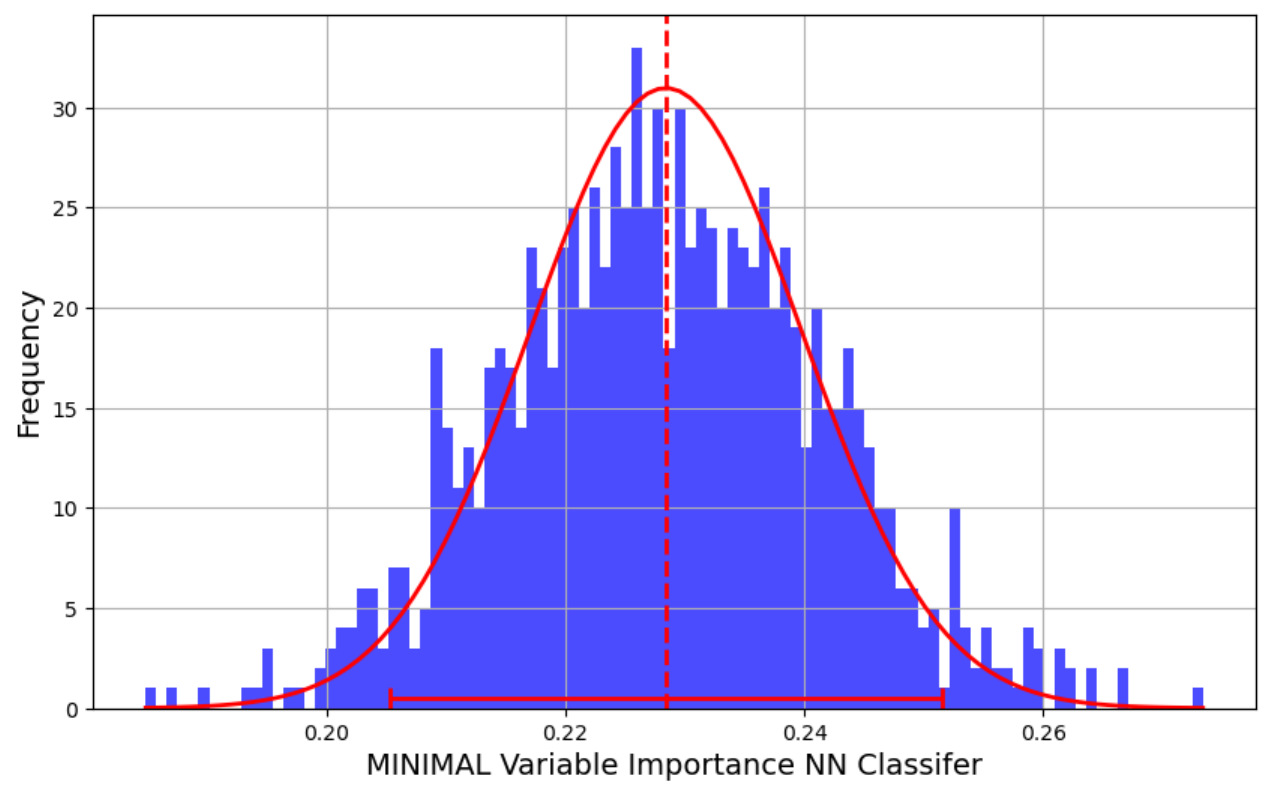}
    \caption{Empirical distribution of 1000 simulated MIMAL values in Simulation 3. The dotted vertical line indicates the true value. The horizontal red line indicates a $95\%$-interval around the truth. }
    \label{fig:nn}
\end{figure}

Let $\sigma(x) \coloneqq e^x(1+e^x)^{-1}$ be the sigmoid function. The simulation runs on a data-generating mechanism taking $M=3$, $n_1=n_2=n_3=700$, and $Y\supm \sim \text{Binomial}\,(700,p\supm)$ with the nonlinear generating process $p\supm = \sigma((X_1^3,X_2^3,X_3^3)\trans\theta\supm)$ where $X\trans \sim \mathcal{U}[-2,2]^3$, and $\theta\supm$ permutes the vector $[0.2,0.6,0.6]$. For ML modeling of $f$, we use a fully connected feed-forward neural network learner with the layer structure set as:
\begin{center}
\begin{tabular}{@{}l@{}}
\texttt{~~~~~nn.Linear(3,6), nn.ReLU(),}\\
\texttt{~~~~~nn.Linear(6,4), nn.ReLU(),}\\
\texttt{~~~~~nn.Linear(4,1), nn.Sigmoid(),}\\
\end{tabular}
\end{center}
which includes totally $57$ parameters, the activation function $\text{ReLU}(x) = \max{(0,x)} \eqqcolon x^+$, and the output layer $\sigma(x) = 1/(1+e^{-x})$. The optimization Algorithm \ref{ttur-gda} has a fixed initialization of the layers. The simulated values center well around the (large-sample-simulated) true saddle point. At most times of the simulations, the learners well-converge to the global saddle point, while in a few cases the algorithm can only find local solutions. The asymptotic unbiasedness and normality of the resulting $\widehat{I}_X$ is well-preserved as shown in Figure \ref{fig:nn} and the CP of MIMAL achieves $94\%$.

\subsection{Simulation 4: Linear Regression Null Model}\label{sec:sim:null}

\begin{figure}[htbp]
    \centering
    \includegraphics[width=0.6\linewidth]{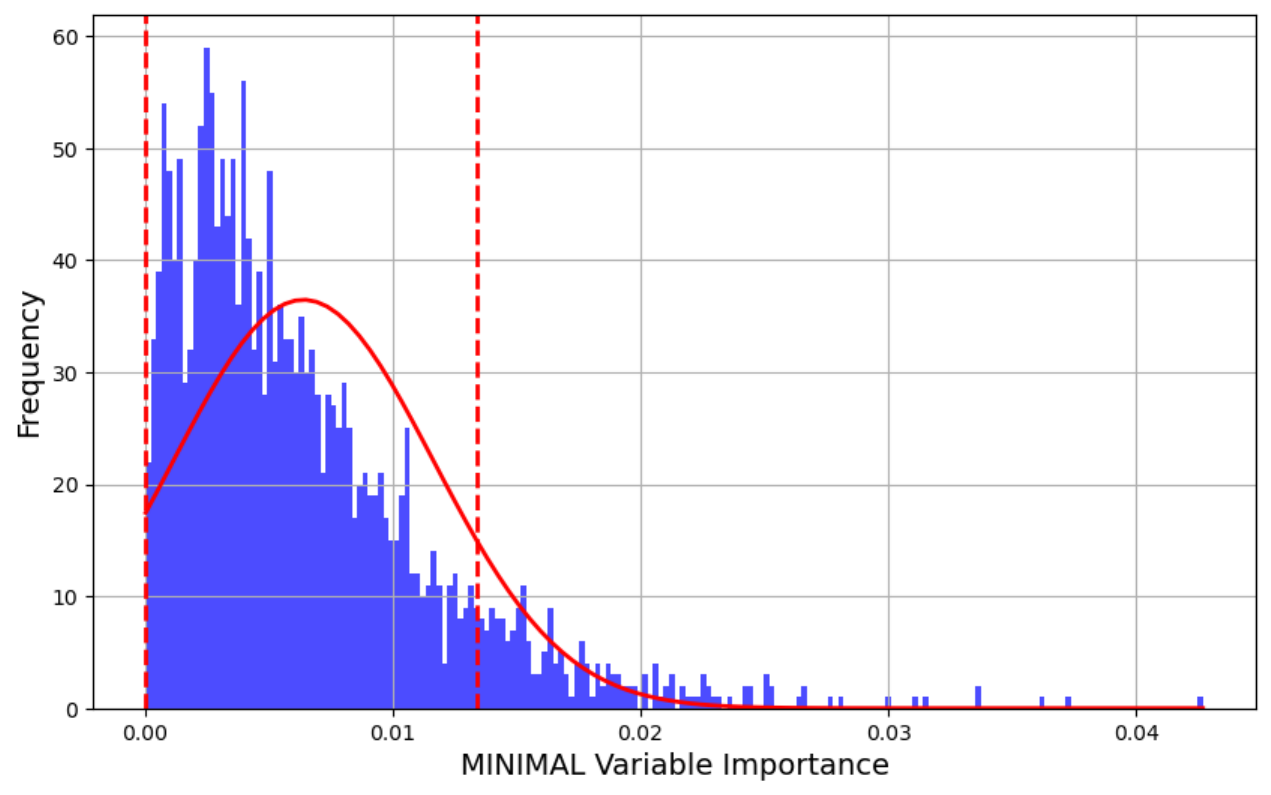}
    \caption{Empirical distribution of 1000 simulated MIMAL values in Simulation 4, against a truncated normal density plot with the simulated sample mean and sample standard deviation. The true $I_X^*=0$ in this case.}
    \label{fig:null}
\end{figure}

We design an experiment where the true population MIMAL objective is zero. We simply consider $M=2$ sources with $n_1=n_2=2000$ and the data generated from $Y\supm = X\trans\theta\supm + Z\trans\gamma\supm + \mathcal{N}(0,1)$ and $(X\trans,Z\trans) \sim \mathcal{U}[-3,3]^5$. In particular, the parameters are 
\begin{align*}
    [\theta^{(1)},\gamma^{(1)}] &= [1,1,1,1] + [0.4,0.3]\\
    [\theta^{(2)},\gamma^{(2)}] &= -\theta^{(1)} + [-0.3,0.2].
\end{align*}
Since $\theta^{(1)}$ and $\theta^{(2)}$ has the same magnitudes and opposite signs, one can show that the linear model maximin solution has first component $\theta^* = [0,0,0,0]$ and the population MIMAL value $I_X^*$ is zero. In this case, as discussed in Remark \ref{rem:2}, the distribution of $\widehat{I}_X$ tends to converge to a non-normal distribution super-efficiently, i.e., faster than $O_p(n^{-1/2})$. This non-normality is demonstrated through the simulated distribution of  $\widehat{I}_X$ in Figure \ref{fig:null}, with $\widehat{\theta}$, $\widehat{\gamma}^{(1)}$, and $\widehat{\gamma}^{(2)}$ being unbiased.

For interval estimate, we employ the variance-inflation methods described in Remark \ref{rem:2} with $\tau=0.1$ or $0.2$. The coverage probability turns out to be $95.9\%$ and $97.9\%$ respectively for $\tau = 0.1$ and $0.2$. An under-coverage of $84.4\%$ happens without any variance inflation i.e. $\tau = 0$, which is expected from the non-normality and super-efficiency of $\widehat{I}_X$ when $I_X^*=0$.

\section{Real-Data Analysis: Beijing Air Pollution}\label{real-data}

In atmospheric science, the term \textit{particulate matter} (PM) stands for microscopic particles of solid or liquid matter suspended in the air. In particular, $\text{PM}_{2.5}$ refers to fine PM that consists of particles in the air that are 2.5 micrometers in diameter or less. In an effort to analyze reduction of Beijing $\text{PM}_{2.5}$ concentration, \cite{Beijing} studied a nonparametric spatial-temporal modelling of Beijing $\text{PM}_{2.5}$ concentration assisted by covariates of meteorological measurements. The data catalogued hourly measurements of $\text{PM}_{2.5}$ from 2013 to 2016 monitored by 12 state-controlled (GuoKong) monitoring sites in Beijing. The measurements are accompanied by 6 meteorological variables. Five of which are numerical: air temperature (TEMP), dew point (DEWP), air pressure (PRES), precipitation (RAIN) and wind speed (WSPM). One of which is categorical: wind direction (WD). The categorical variable WD is transformed to a four dimensional indicator vector according to the four directions $[\text{N},\text{E},\text{S},\text{W}]$, for instance $\text{NE}\mapsto [1,1,0,0]\in \{0,1\}^4$. We then concatenate WD and WSPM into a single group of factors to represent the wind condition (WC). 

All variables are collected in $M=3$ geographically distant monitoring sites including Aoti, Changping and Shunyi. The data has a natural paired structure as introduced in Remark \ref{rem:3} that all the hourly measurements are aligned according to the time point across the three sources. We randomly sample $n_1=n_2=n_3=700$ observations (time points) among the four months from November, 2013 to February, 2014. We conduct two MIMAL analyses separately using parametric models and KRR, to infer the variable importance of each covariate $X$ on the $\text{PM}_{2.5}$ outcome $Y$ in adjustment for the remaining covariates $Z$, in a similar spirit with the LOCO strategy. Since $Y$ is continuous, we take $\ell(y,u)=-(y-u)^2$. For parametric regression, we form the interaction basis $\boldsymbol{X}_{\text{inter}} =[XZ_1,\ldots,XZ_{{\rm dim}(Z)}]$ and set
\[
f(X,Z) = \theta_0 X + \boldsymbol{X}_{\text{inter}}\trans\theta;\quad  g\supm(Z)= \gamma_0\supm+Z\trans\gamma\supm.
\]
For KRR, we use the RBF kernel as in Section \ref{simul2} with $\sigma=0.2$, and the penalty coefficient set as $1/n_1$. In addition, to ensure training stability, we introduce a small ridge regularization on $q$ as described in Remark \ref{rem:non-uni} with the penalty coefficient $\delta=0.001$. 

\begin{figure}[H]
    \centering
    \subfigure{
        \includegraphics[width=0.47925\textwidth]{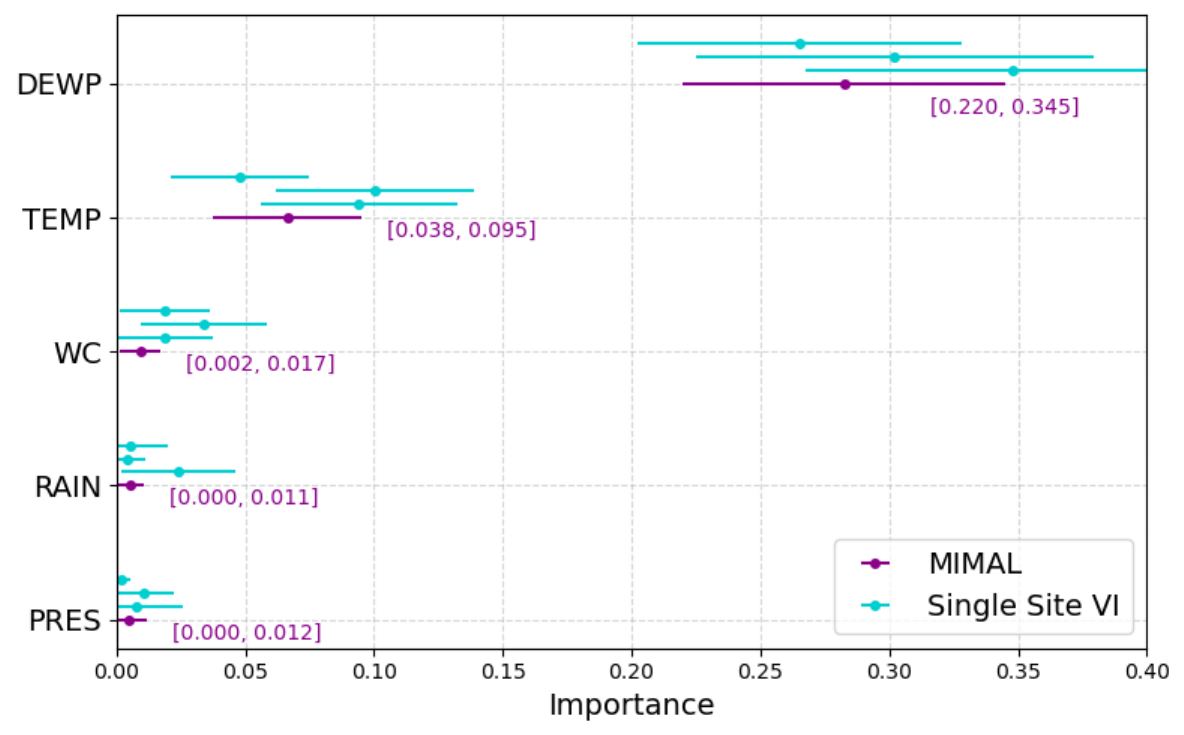}
    }
    \hfill
    \subfigure{
        \includegraphics[width=0.47925\textwidth]{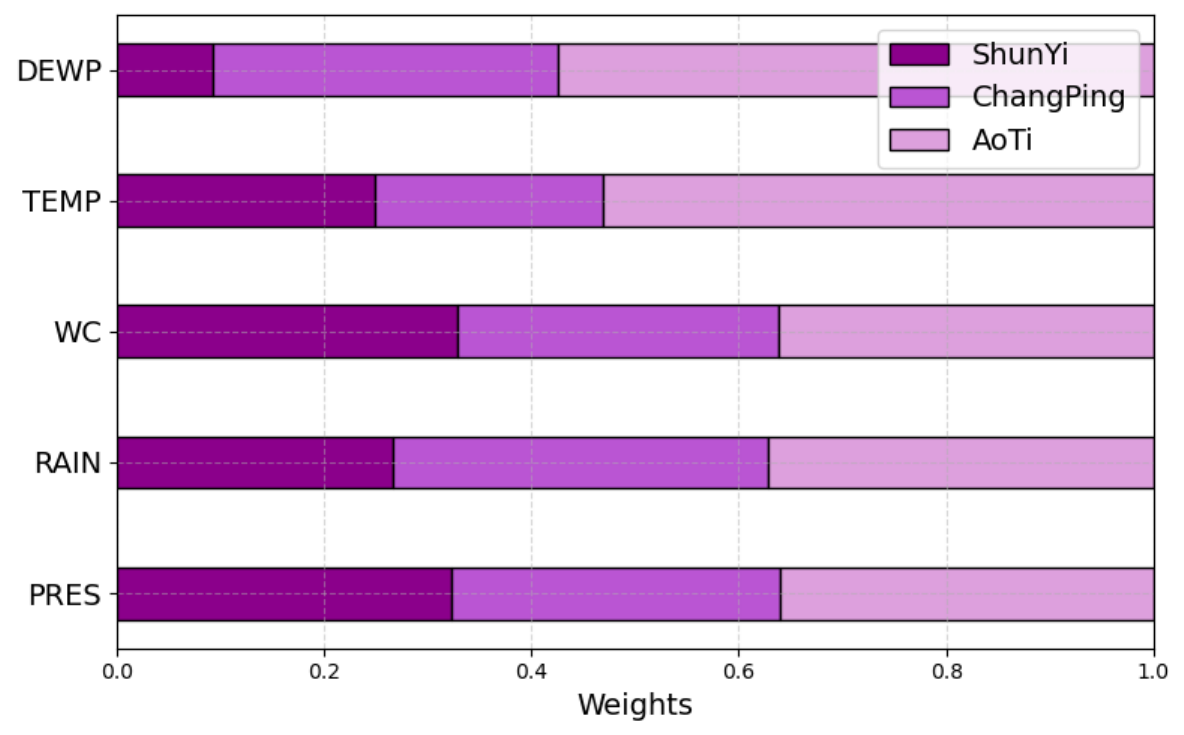}
    }
    \caption{\label{Beijing-GLM} \textit{MIMAL with parametric regression.} Left figure: 95\% CIs of the MIMAL variable importance for every predictor learned by parametric regression. Right figure: Fitted $q$-component in the saddle point.}
\end{figure}

\begin{figure}[H]
    \centering
    \subfigure{
        \includegraphics[width=0.47925\textwidth]{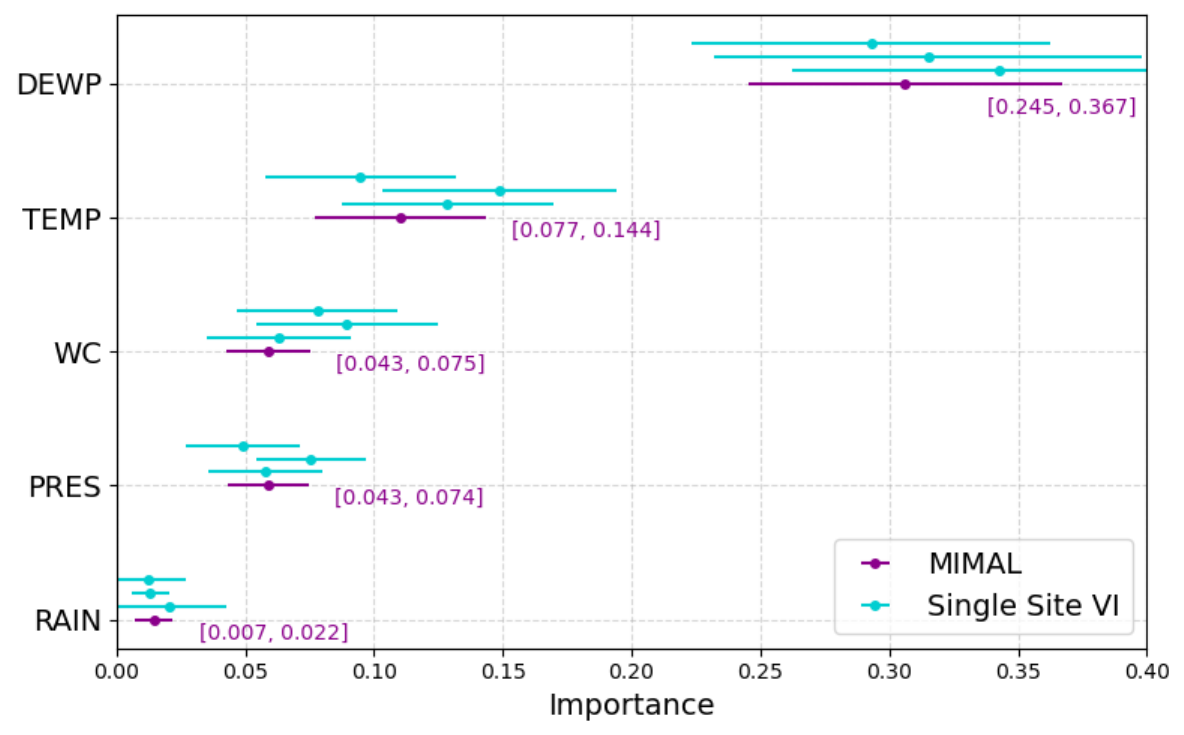}
    }
    \hfill
    \subfigure{
        \includegraphics[width=0.47925\textwidth]{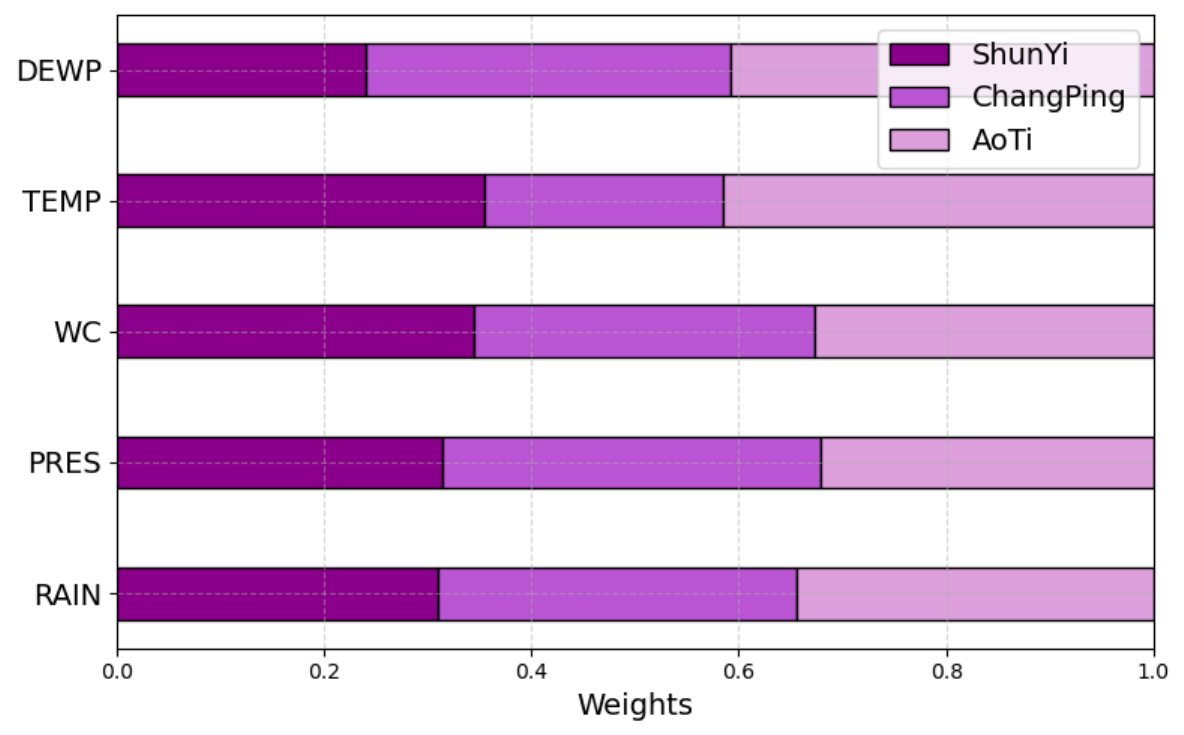}
    }
    \caption{\label{Beijing-KRR} \textit{MIMAL with KRR.} Left figure: 95\% CIs of the MIMAL variable importance for every predictor learned by KRR. Right figure: Fitted $q$-component in the saddle point.}
\end{figure}

The resulting 95\% CIs for $I_X^*$ and each source-specific variable importance $I\supm_X$ as well as the fitted $\widehat{q}$ for the predictors are presented in Figures \ref{Beijing-GLM} and \ref{Beijing-KRR}, respectively for the parametric and KRR constructions. DEWP stands out as the most important predictor for $\text{PM}_{2.5}$ across the three sources, with an $[0.22, 0.35]$ CI for $I_X^*$ using parametric regression and an $[0.25, 0.37]$ CI using KRR. This agrees with the conclusion in recent scientific literature \citep[e.g.]{pm2.5_review} that humidity (dew point) has the most dominant effect on $\text{PM}_{2.5}$ particle formation in the atmosphere. Air temperature (TEMP) and the group of wind condition variables (WC) also have their importance variable on $\text{PM}_{2.5}$ significantly larger than $0$ with both learning methods. Differently, precipitation (RAIN) and air pressure (PRES) only show a moderately significant $I^*_X$ when using KRR but have their CIs covering $0$ with parametric regressions.

Due to the small ridge penalty on $q$, the fitted $I^*_X$'s are not rigorously smaller than the smallest $I\supm_X$ as its original population version is supposed to be. Nevertheless, we still observe that MIMAL produces a stable summary of the variable importance across multiple sources, as discussed in Section \ref{sec:framework}. In addition, the variable importance estimated using KRR is moderately larger in values than their parametric model counterparts. This is because KRR is more capable of capturing non-linear relationships and could explain more variance of $Y$ with $X$.

\section{Conclusion and Discussion}\label{sec:disc}
In this paper, we introduced MIMAL, a novel framework for assessing stable variable importance across heterogeneous data sources via adversarial learning. By optimizing the worst-case predictive reward among the sources, MIMAL quantifies variable importance in a manner that emphasizes generalizability and stability. The framework incorporates advanced ML techniques and ensures asymptotic unbiasedness and normality of $\widehat{I}_X$, even under complex data and model settings. Extensive simulations and real-world application highlight its utility in diverse scenarios, showcasing its capacity to address challenges in multi-source variable importance analysis with interpretability and reliability. 

We now discuss several future directions to improve and generalize our work. Firstly, it is potentially interesting to extend our framework to a transfer learning setting with covariate shift. In this scenario, we are interested in quantifying the predictive importance of $X$ on some target population $\mathbb{Q}_{X,Z}\neq\mathbb{P}_{X,Z}\supm$'s. Meanwhile, on the target, we only have samples of $(X,Z)$  without any observations of $Y$, which calls for knowledge transfer of the outcome models $\mathbb{P}_{Y\mid X,Z}\supm$'s from the sources, known as the covariate shift issue \citep[e.g.]{gretton2009covariate}. For stable variable importance measure in this setting, one could use importance weighting that corrects for the covariate shift by re-weighting the sample on each source $m$ with the density ratio between $\mathbb{Q}_{X,Z}$ and $\mathbb{P}_{X,Z}\supm$. Furthermore, the doubly robust framework of \cite{liu2023augmented} can be potentially incorporated to provide more robust and efficient inference. 

Secondly, as discussed in Remark \ref{rem:non-uni}, the strict convexity Assumption \ref{restriced_convexity} is made to ensure the uniqueness of $\bar{q}$, which is necessary for the normality of $\widehat{I}_X$ and commonly used in group DRoL literature \citep[e.g.]{Zijian_groupDRO}. Adding a ridge penalty on $q$ as in (\ref{equ:ridge}) is a convenient way to fix this issue in practice but incurs an undesirable change to the objective function. A future direction is to maintain valid inference with potentially non-normal $\widehat{I}_X$ obtained from the original objective without any regularization on $q$. Finally, to make our framework more general and flexible, it is desirable to incorporate other non-gradient-based ML methods like random forest and $k$-nearest neighbors.

\section*{Acknowledgement}
The authors would like to thank Bharath Sriperumbudur (Penn State University) for helpful discussion on RKHS.

\bibliographystyle{apalike}
\bibliography{ref.bib}

\end{document}